\documentclass[journal]{IEEEtran}

\usepackage{pgfplots}
\usepackage[bookmarks,colorlinks]{hyperref}
\usepackage[linesnumbered,ruled,lined]{algorithm2e}
\usepackage{enumitem}
\usepackage{algpseudocode}
\usetikzlibrary{shapes.multipart,intersections}
\usepackage{cite}
\usepackage{amsmath,amssymb,amsfonts,amsthm,steinmetz}
\usepackage{graphicx}
\usepackage{mathrsfs}  
\usepackage{textcomp}
\usepackage{acronym}
\usepackage{xcolor}
\usepackage{upgreek,xspace}

\usepackage{comment}

\usepackage{tikz}
\usetikzlibrary{calc}
\makeatletter
\newcommand{\gettikzxy}[3]{%
  \tikz@scan@one@point\pgfutil@firstofone#1\relax
  \edef#2{\the\pgf@x}%
  \edef#3{\the\pgf@y}%
}
\makeatother

\usepackage[draft]{todonotes}

\usepackage{esvect}

\usepackage{booktabs}

\usepackage{times}
\usepackage{bm}
\usepackage{amsmath}
\usepackage{amssymb}
\usepackage{stmaryrd}
\usepackage{babel}
\usepackage{graphics, graphicx}
\usepackage{xcolor}
\usepackage{gensymb}
\usepackage{cite}
\usepackage{enumitem}
\usepackage{url}
\newtheorem{rem}{Remark}

\begin{document}

\title{Benefits of Mutual Coupling in \\Dynamic Metasurface Antennas}

\author{Hugo~Prod'homme,~Jean~Tapie,~Luc~Le~Magoarou,~\IEEEmembership{Member,~IEEE},~and~Philipp~del~Hougne,~\IEEEmembership{Member,~IEEE}
\thanks{
H.~Prod'homme, J.~Tapie, and P.~del~Hougne are with Univ Rennes, CNRS, IETR - UMR 6164, F-35000, Rennes, France (e-mail: \{hugo.prodhomme; jean.tapie; philipp.del-hougne\}@univ-rennes.fr).
}
\thanks{
L.~Le~Magoarou is with INSA Rennes, CNRS, IETR - UMR 6164, F-35000, Rennes, France (e-mail: luc.le-magoarou@insa-rennes.fr).
}
\thanks{\textit{(Corresponding Author: Philipp del Hougne.)}}
\thanks{H.~Prod'homme and J.~Tapie contributed equally to this work.}
\thanks{This work was supported in part by the ANR France 2030 program (project ANR-22-PEFT-0005), the ANR PRCI program (project ANR-22-CE93-0010), CREACH LABS (project AdverPhy), the European Union's European Regional Development Fund, and the French region of Brittany and Rennes Métropole through the contrats de plan État-Région program (projects ``SOPHIE/STIC \& Ondes'' and ``CyMoCoD'').}
}

\maketitle

\begin{abstract}
Dynamic metasurface antennas (DMAs) are a promising embodiment of next-generation reconfigurable antenna technology to realize base stations and access points with reduced cost and power consumption. A DMA is a thin structure patterned on its front with reconfigurable radiating metamaterial elements (meta-atoms) that are excited by waveguides or cavities. Mutual coupling (MC) between the meta-atoms can result in a strongly non-linear dependence of the DMA's radiation pattern on the configuration of its meta-atoms. 
However, besides the obvious algorithmic challenges of working with physics-compliant DMA models, it remains unclear how MC in DMAs influences the ability to achieve a desired wireless functionality. 
In this paper, we provide theoretical, numerical and experimental evidence that strong MC in DMAs increases the radiation pattern sensitivity to the DMA configuration and thereby boosts the available control over the radiation pattern, improving the ability to tailor the radiation pattern to the requirements of a desired wireless functionality.
Counterintuitively, we hence encourage next-generation DMA implementations to enhance (rather than suppress) MC, in combination with suitable physics-compliant modeling and optimization.
We expect the unveiled mechanism by which MC boosts the radiation pattern control to also apply to other reconfigurable antenna systems based on tunable lumped elements. 
\end{abstract}


\section{Introduction}\label{sec_Introduction}

Multi-antenna systems have become increasingly important in wireless communications since the late 1990s~\cite{foschini1998limits,telatar1999capacity} and play a critical role in recent and next-generation wireless communications networks. In particular, multi-antenna systems enable multiple-input multiple-output (MIMO) techniques that leverage spatial (rather than spectral or temporal) resources to boost key metrics like the capacity and spectral efficiency of wireless systems. Traditionally, multi-antenna systems comprise a \textit{static} array of antennas, each with an individual radio-frequency (RF) chain. By optimizing the coherent signal that excites the antennas, multi-antenna systems enable dynamic beam-steering to a single or multiple users, as well as many more advanced wireless functionalities. 

However, the scalability of such multi-antenna systems is fundamentally limited by considerations of cost, power consumption, weight, and footprint. To overcome these limitations, next-generation \textit{reconfigurable} antenna technology seeks to exploit additional degrees of freedom in the physical layer. 
The software-controllable electromagnetic properties of the resulting hybrid analog/digital architectures enable the dynamic reconfiguration of radiation patterns. Practical implementations are typically based on tunable lumped elements offering electrical tunability at spatially discrete locations. 
Early examples are \textit{electronically steerable passive array radiator (ESPAR) antennas}~\cite{harrington1978reactively,schlub2003seven,sun2004fast,kawakami2005electrically,lu2005dielectric,luther2012microstrip,movahedinia2018}, where a few driven antennas are surrounded by tunably loaded parasitic elements.
Another important example are 
\textit{reconfigurable pixel-based antennas}; these are grids of metallic patches interconnected by tunable lumped elements~\cite{Flaviis_PixelAntenna,rodrigo2012frequency,MURCH_TAP_PixelAntenna}, where some of the patches are the main radiators and the tunable lumped elements control the coupling between all patches. 
A third example are \textit{dynamic metasurface antennas (DMAs)}; these are surfaces patterned with radiating reconfigurable meta-atoms that are excited by waveguides or cavities~\cite{sleasman2015dynamic,yurduseven2018dynamically,sleasman2020implementation,del2020learned,boyarsky2021electronically,shlezinger2021dynamic,jabbar202460,yven2025end}. In DMAs, the tunable meta-atoms are simultaneously the tuning mechanism and the radiation mechanism. We focus for concreteness on DMAs in this paper; however, we expect the fundamental principles that we unveil to apply more broadly to any reconfigurable antenna based on tunable lumped elements.

In optimizing the software-controllable configuration of a DMA for a desired wireless functionality, two aspects of the \textit{dependence of the DMA radiation pattern on the DMA configuration} matter for the wireless practitioner:
\begin{enumerate}
    \item \textit{Strength.} A strong dependence is desirable to maximize the ability to tailor the DMA radiation pattern to a desired wireless functionality.
    \item \textit{Simplicity.} A mathematically simple dependence is preferable to facilitate the algorithmic design.
\end{enumerate}
It is relatively well understood that mutual coupling (MC) between the reconfigurable meta-atoms\footnote{The question of how MC between physically tunable elements impacts the influence of these physical-layer degrees-of-freedom on system performance is clearly different from the well-studied impact of MC in static arrays on the performance of systems for which only the coherent excitation signals are adjustable~\cite{Wallace_Jensen_MC_2004,Neil_MC_2018}.} profoundly influences the mathematical nature of the mapping from DMA configuration to DMA radiation pattern. Specifically, this mapping is generally \textit{non-linear} due to MC, as seen in physics-compliant models for various DMA hardware designs~\cite{pulido2018analytical,yoo2019analytic,yoo2020analytic,williams2022electromagnetic}. Because the mapping's non-linearity severely complicates the algorithmic design, MC is generally considered a vexing nuance. Wireless practitioners hence either avoid these complications by neglecting or deliberately minimizing MC in DMAs to achieve an effectively affine mapping~\cite{smith2017analysis,yurduseven2018dynamically,shlezinger2019dynamic,wang2020dynamic,shlezinger2021dynamic,boyarsky2021electronically,RISnDMA_2024,DMA2023,jabbar202460,DMA_WPT}, or they develop advanced algorithms based on physics-compliant models that capture MC in DMAs~\cite{del2020learned,qian2022noise,gavras2024circuit}. However, the influence of MC on the strength of the dependence of the DMA radiation pattern on the DMA configuration is generally not considered.

Recently, inspired by earlier work on the benefits of multi-path propagation in wireless localization~\cite{del2021deeply}, we presented preliminary full-wave numerical evidence that MC boosts the influence of the DMA configuration on the DMA radiation pattern~\cite{prod2024mutual}; we further suggested that stronger MC should therefore improve the ability to tailor the DMA radiation pattern to a desired wireless functionality~\cite{prod2024mutual}. This generally overlooked effect of MC in DMAs can drastically change the assessment of the role of MC in DMAs. In essence, a \textit{trade-off between strength and simplicity} of the dependence of the DMA radiation pattern on the DMA configuration emerges. Intuitively, these two effects of MC can be understood in terms of multi-bounce paths between the meta-atoms\footnote{The relation between physics-compliant models and a multi-bounce-path picture has been worked out in the context of RIS-parametrized radio environments using a coupled-dipole formalism in~\cite{rabault2024} and using multi-port network theory in Sec.~II.A of~\cite{del2024physics}.}: The stronger the MC in the DMA is, the more the wave bounces back and forth between meta-atoms, resulting in an accumulation of more sensitivity to the DMA configuration, which in turn yields a \textit{i) }stronger and \textit{ii)} more strongly non-linear (i.e., less simple) dependence of the DMA radiation pattern on the DMA configuration. 
Consequently, it may be desirable to deploy DMA hardware with enhanced (rather than suppressed) MC in combination with suitable physics-compliant models and optimization techniques.

Motivated by the above, this paper presents a detailed theoretical, numerical and experimental analysis of the relation between the following four pivotal quantities:
\begin{enumerate}
    \item MC strength between the DMA's meta-atoms,
    \item sensitivity of the DMA radiation pattern to the DMA configuration, 
    \item ability to tailor the DMA radiation pattern to the needs of a desired wireless functionality, and
    \item linearity of the mapping from DMA configuration to DMA radiation pattern.
\end{enumerate}
The main contributions are summarized as follows:
\begin{itemize}
    \item We formulate a compact, generic (implementation-agnostic), physics-compliant DMA model within the coupled-dipole framework. 
    Our formulation collapses architecture specifics into a background coupling operator, so that our formulation is architecture-agnostic and maximally compact.\footnote{A related, similarly compact formulation in terms of impedance parameters was recently proposed in~\cite{almunif2025network}; the first preprint of~\cite{almunif2025network} appeared shortly after the first preprint of our present manuscript.} Our formulation transparently exposes the mathematical role of MC in DMAs. We also clarify how to recover common simplified linear models upon neglecting MC. We further expect our formulation's compactness and generality to be ideal for future efforts to experimentally estimate the model parameters (see discussion in Sec.~\ref{sec_Discussion}). Our formulation complements (rather than supersedes) existing architecture-specific coupled-dipole formulations like~\cite{yoo2019analytic,yoo2020analytic} which remain indispensable to construct the background coupling operator in numerical studies.
    \item We derive analytical expressions and upper bounds for the normalized Jacobian and Hessian of the DMA radiation pattern with respect to the DMA configuration. The upper bounds monotonically increase with the MC strength.
    \item We systematically examine in full-wave simulations how the four pivotal quantities listed above depend on the MC strength in a DMA. We reveal clear trends of positive correlation between MC strength and sensitivity, achieved channel gain enhancement, and non-linearity of the mapping from configuration to radiation pattern.
    \item We conduct experiments with a DMA prototype operating in the lower K-band with 1-bit programmable meta-atoms. We introduce an original time-gating technique to adjust the effective MC strength in post-processing. We report clear experimental evidence for the numerically observed trends.
\end{itemize}

\textit{Organization.} In Sec.~\ref{sec_system_model}, we present a compact,  implementation-agnostic, physics-compliant system model for DMAs.
In Sec.~\ref{sec_Theory}, we analytically derive upper bounds on the normalized Jacobian and Hessian of the DMA radiation pattern with respect to the DMA configuration, as a function of the DMA's MC strength.
In Sec.~\ref{sec_numerical}, we provide full-wave numerical evidence for the trade-off between strength and simplicity of the mapping from configuration to radiation pattern, and we demonstrate how a stronger sensitivity improves the ability to optimize the channel gain with a DMA. 
In Sec.~\ref{sec_experimental}, we use a DMA prototype to experimentally confirm these trends.
We close in Sec.~\ref{sec_Conclusion} with a conclusion and a brief outlook to future work.

\textit{Notation.} $\mathbf{A} = \mathrm{diag}(\mathbf{a})$ denotes the diagonal matrix $\mathbf{A}$ whose diagonal entries are $\mathbf{a}$. $\mathbf{b} = \mathrm{diag}(\mathbf{B})$ denotes the vector $\mathbf{b}$ containing the diagonal entries of the matrix $\mathbf{B}$. 
$\left[ \mathbf{A}^{-1} \right]_\mathcal{BC}$ denotes the block of $\mathbf{A}^{-1}$ selected by the sets of indices $\mathcal{B}$ and $\mathcal{C}$. $\left[ \mathbf{a} \right]_\mathcal{B}$ denotes the vector formed by the elements of $\mathbf{a}$ selected by $\mathcal{B}$.
$\mathbf{a}^\top$ denotes the transpose of $\mathbf{a}$. 
$\mathbf{I}_a$ denotes the $a \times a$ identity matrix. $\mathbf{0}_a$ denotes the $a \times a$ zero matrix.
$\mathbf{0}_\mathcal{BC}$ denotes the $|\mathcal{B}| \times |\mathcal{C}|$ zero matrix, with $|\mathcal{B}|$ denoting the cardinality of $\mathcal{B}$.

\section{System Model}
\label{sec_system_model}

In this section, we develop a physics-compliant DMA system model based on a coupled-dipole formalism that only explicitly describes the primary entities (feeds and meta-atoms) and lumps the influence of all other entities into the coupling coefficients between the primary entities. Thereby, we formulate a physics-compliant DMA model in the most compact and generic (implementation-agnostic) manner, allowing us to derive general analytical insights about strength and simplicity of the dependence of the DMA radiation pattern on the DMA configuration in Sec.~\ref{sec_Theory}.

\begin{figure}
    \centering
    \includegraphics[width=0.8\columnwidth]{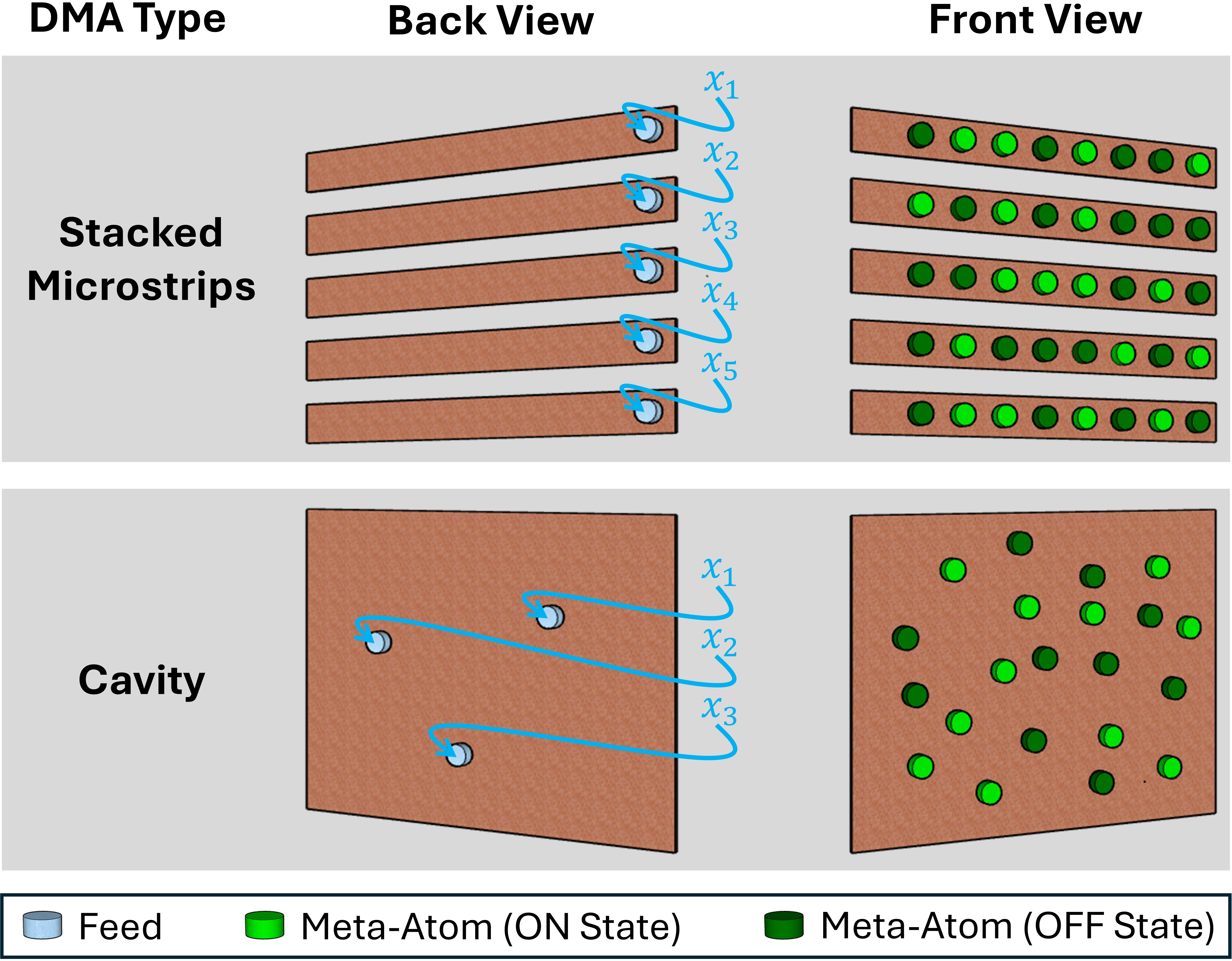}
    \caption{A DMA is a thin structure patterned with $N_\mathrm{M}$ programmable meta-atoms (green) on the front and $N_\mathrm{F}$ feeds (light blue) on the back. During transmission, the input signal $\mathbf{x} = [x_1, x_2, ..., x_{N_\mathrm{F}}]$ excites the DMA via the feeds. The DMA architecture and the configuration of the meta-atoms determine how the signal couples from the feeds to the meta-atoms. The meta-atoms radiate the signal according to their coupling to the feeds and their configuration. 
    The coupling between feeds and meta-atoms is mediated either via microstrips (top row) or the cavity (bottom row). In the former case, any given feed only couples to meta-atoms that are on the same microstrip.}
    \label{Fig1}
\end{figure}

At a high level of abstraction, any DMA can be described as consisting of $N_\mathrm{F}$ feeds and $N_\mathrm{M}$ meta-atoms whose electromagnetic coupling is mediated by waveguides or cavities in a way that is implementation-specific (see Fig.~\ref{Fig1}). During transmission, the input signal excites the feeds and ultimately the meta-atoms because they are coupled to the feeds; then, the meta-atoms radiate the signal to the scene. During reception, the reverse happens.

Existing physics-compliant DMA models (e.g.,~\cite{pulido2017polarizability,pulido2018analytical,yoo2019analytic,yoo2020analytic,williams2022electromagnetic}) seek to explicitly describe each detail of a given DMA architecture in closed form to determine the coupling between feeds and meta-atoms. Thereby, they enable numerical studies such as ours in Sec.~\ref{sec_numerical}. However, thereby they are also inevitably specific to a given DMA architecture, their experimental applicability is vulnerable to fabrication inaccuracies, and high-level conceptual insights are somewhat obscured (such as the fundamental trade-off between strength and simplicity of the dependence of the radiation pattern on the configuration that we investigate here). 

In this section, we seek to distill the essential wave physics without getting into the weeds of deriving a closed-form expression for analytically evaluating this coupling in a given DMA hardware. We hence do \textit{not} make any assumption about the DMA hardware specifics such that our compact model applies to both types of common DMA implementations shown in Fig.~\ref{Fig1}. 
We only assume that the DMA's static parts are linear, passive, time-invariant and reciprocal, which is generally the case, and that the DMA's tunable parts (i.e., the meta-atoms) are very small compared to the operating wavelength, which is justified by their intrinsic sub-wavelength nature and an accurate approximation according to earlier work~\cite{pulido2017polarizability,pulido2018analytical,yoo2019analytic,yoo2020analytic}.  
Throughout this paper, we do not explicitly print frequency dependence because we study DMAs in single-frequency operation.

\subsection{Wave Propagation in the DMA}

Our starting point is a description of the $N = N_\mathrm{F} + N_\mathrm{M}$ primary entities (feeds and meta-atoms) as dipoles. The validity of this approach was validated in commercial simulation software and experiments in~\cite{pulido2017polarizability,pulido2018analytical,yoo2019analytic,yoo2020analytic}.
Specifically, one dipole is assigned to the location of each primary entity. The $i$th dipole is a point-like polarizable particle with a specific orientation and of infinitesimal size $\delta_i$ that accumulates a charge separation of $q_i$ and $-q_i$ across its ends when an electromagnetic field $e_i$ is applied to it along its orientation. This charge separation gives rise to a dipole moment $p_i = q_i \delta_i$. The tendency of a dipole to build up a dipole moment under the application of an electromagnetic field  is characterized by its polarizability: $\alpha_i = p_i / e_i$.
The polarizability of the feeds is fixed whereas the polarizability of the meta-atoms is tunable. We define the DMA configuration as  
\begin{equation}
    \mathbf{c} =  \left[ \alpha_0^{-1} - \alpha_i^{-1} \  \vert \  i \in \mathcal{M} \right] \in \mathbb{C}^{N_\mathrm{M}},
\end{equation}
where $\alpha_0$ is a reference polarizability value; $\mathcal{M}$ [$\mathcal{F}$] denotes the set of dipole indices associated with meta-atoms [feeds]. 
Dipoles that have accumulated dipole moments reradiate electromagnetic fields at the same frequency as the initial impinging field. 
The field at the location and along the orientation of the $j$th dipole originating from a unit dipole moment at the $i$th dipole is defined as the background Green's function $G_{ji}$. The term ``background'' highlights that $G_{ji}$ is generally not a free-space Green's function but specific to the DMA hardware involving waveguides or cavities. 
The field $e_i$ at the location and along the orientation of the $i$th dipole is hence the superposition of the initial impinging field and the reradiated secondary fields:
\begin{equation}
   e_i = \alpha_i^{-1} p_i =  e_i^\mathrm{inc} + \sum_{j=1}^N G_{ij} p_j.
\end{equation}
We can now self-consistently solve this system of coupled equations for the dipole moments~\cite{PhysFad}:
\begin{equation}
    \mathbf{p} = \mathbf{W}^{-1} \mathbf{e}^\mathrm{inc},
   \label{eq_3}
\end{equation}
where $\mathbf{p} = [p_1,  \dots, p_N] \in \mathbb{C}^{N}$, $\mathbf{e}^\mathrm{inc} = [e_1^\mathrm{inc}, \dots, e_N^\mathrm{inc}] \in \mathbb{C}^{N}$ with $\left[ e^\mathrm{inc}_i \  \vert \  i \in \mathcal{F} \right] \in \mathbb{C}^{N_\mathrm{F}}$ proportional to the input signal $\mathbf{x}\in\mathbb{C}^{N_\mathrm{F}}$ and $e_i^\mathrm{inc} = 0 \ \forall \ i \in \mathcal{M}$, and\footnote{Generally, $G_{ii}\neq 0$ for ``background'' Green's functions.} 
\begin{equation}
W_{i,j}=\begin{cases}
\alpha_{i}^{-1} - G_{ii}, & i=j\\
-G_{ij}, & i\neq j
\end{cases}.
\end{equation}
To highlight the influence of the DMA configuration $\mathbf{c}$, the following partition of $\mathbf{W}$ is insightful:
\begin{equation}
    \mathbf{W}(\mathbf{c})  = \begin{bmatrix} 
	\mathbf{W}^0_{\mathcal{FF}} & \mathbf{W}^0_{\mathcal{FM}} \\
	\mathbf{W}^0_{\mathcal{MF}} & \mathbf{W}^0_{\mathcal{MM}} - \mathrm{diag}(\mathbf{c})\\
	\end{bmatrix},
    \label{eq5}
\end{equation}
where 
\begin{equation}
W^0_{i,j}=\begin{cases}
\alpha_{i}^{-1} - G_{ii}, & i=j \in \mathcal{F}\\
\alpha_{0}^{-1} - G_{ii}, & i=j \in \mathcal{M}\\
-G_{ij}, & i\neq j
\end{cases}.
\end{equation}
At this stage, the following clarifying remarks can be helpful:
\begin{rem}
    We use scalar notation for $G_{ij}$ because our careful definition implies that it is the projection of the dyadic Green's tensor onto the orientations of the $i$th and $j$th dipoles. 
\end{rem}
\begin{rem}
    In contrast to decades-old discrete-dipole approximations (DDAs) of light scattering~\cite{purcell1973scattering,mulholland1994light,de1998point} and metamaterials~\cite{bowen2012using} that discretize an entire system into dipoles, we only assign dipoles to the primary entities. Meanwhile, the scattering by all other entities (in particular, waveguides or cavities) is lumped into the background Green's functions. Thereby, we obtain the most compact and generic model formulation.\footnote{Similar compact physics-compliant models exist for wireless channels parametrized by reconfigurable intelligent surfaces (RISs)~\cite{prod2023efficient,sol2023experimentally,sol2024optimal,tapie2023systematic,del2024minimal,V2NA_2p0,del2024physics}.}
\end{rem}
\begin{rem}
    The matrix inversion in Eq.~(\ref{eq_3}) can be interpreted with a multi-bounce-path picture by casting it into an infinite series~\cite{rabault2024}. It causes the non-linearity in the mapping from DMA configuration to DMA radiation pattern (see below).
\end{rem}

\subsection{DMA Radiation Pattern}

Having determined the meta-atoms' dipole moments via Eq.~(\ref{eq_3}), we now derive the DMA's radiation pattern. 
As explained in~\cite{yoo2020analytic} (see discussion around (21) and (22) therein) as well as~\cite{yoo2019modeling}, the radiated field can be evaluated by treating the meta-atoms as dipoles in free space, upon scaling the dipole moments found in (\ref{eq_3}) by a real-valued scalar $\nu$. This scaling accounts, among other effects, for the electromagnetic images of the dipoles. For the analysis in this paper, the value of this scaling factor is immaterial.
The radiated field has three polarization components ($x$, $y$ and $z$ in Cartesian coordinates); the $k$-component (where $k\in\{x,y,z\}$) of the radiated field at some position $\mathbf{r}$ in the scene is the superposition of the $k$-components of the fields radiated by the meta-atoms to that position:\footnote{Implicit in Eq.~(\ref{eq7}) is that the meta-atoms radiate toward the scene like dipoles in free space (which is a reasonable assumption except for grazing angles) and that there is no static direct contribution from the feeds to the field in the scene (which is reasonable because the feeds are embedded within the DMA rather than patterned onto its front surface like the meta-atoms).}
\begin{equation}
\begin{split}
    e^{\mathrm{rad},k} (\mathbf{r},\mathbf{c}) &= \nu\sum_{i =1}^{N_\mathrm{M}} G_k^\mathrm{FS}(\mathbf{r},\mathbf{r}_{\mathcal{M}_i}) p_{\mathcal{M}_i} (\mathbf{c}) = \left( \mathbf{v}_k(\mathbf{r})\right)^\top \mathbf{p}_\mathcal{M}(\mathbf{c}) \\&= \left( \mathbf{v}_k(\mathbf{r})\right)^\top  \left[ \left(\mathbf{W} (\mathbf{c}) \right)^{-1} \mathbf{e}^\mathrm{inc} \right]_\mathcal{M} \\ &= \left( \mathbf{v}_k(\mathbf{r})\right)^\top \left[ \left(\mathbf{W} (\mathbf{c}) \right)^{-1} \right]_\mathcal{MF} \mathbf{e}^\mathrm{inc}_\mathcal{F},
    \end{split}
    \label{eq7}
\end{equation}
where $\mathbf{r}_{\mathcal{M}_i}$ is the location of the $i$th meta-atom and $G_k^\mathrm{FS}(\mathbf{r},\mathbf{r}_{\mathcal{M}_i}) $ is the projection of the \textit{free-space} dyadic Green's tensor between positions $\mathbf{r}_{\mathcal{M}_i}$ and $\mathbf{r}$ onto the field component $k$ at position $\mathbf{r}$ and the field component parallel to the orientation of the $i$th dipole at position $\mathbf{r}_{\mathcal{M}_i}$ (see Eq.~(9.18) in~\cite{jackson1999electrodynamics} for the closed-form expression), and $\mathbf{v}_k(\mathbf{r}) = \nu \left[G_k^\mathrm{FS}(\mathbf{r},\mathbf{r}_i)  \  \vert \  i \in \mathcal{M} \right] \in \mathbb{C}^{N_\mathrm{M}}$. The radiation pattern is hence a linear projection of the meta-atoms' dipole moments to the scene. 

\begin{rem}
We note a qualitative difference between physics-compliant models of DMAs and RIS-parametrized radio environments. Whereas the $\mathcal{MF}$ block of $\mathbf{W}^{-1}$ (i.e., $\left[ \mathbf{W}^{-1} \right]_\mathcal{MF}$) matters in Eq.~(\ref{eq7}) for the DMA case, the channel matrix is given by an off-diagonal block of the $\mathcal{FF}$ block in the RIS case (substituting feeds with antennas and DMA meta-atoms with RIS elements)~\cite{PhysFad,prod2023efficient,rabault2024,sol2023experimentally}.
\end{rem}

\begin{rem}
    An equivalent physics-compliant DMA model can also be formulated based on circuit theory as opposed to the coupled-dipole formalism; for instance,~\cite{williams2022electromagnetic} presents a circuit-theoretic formulation specific to microstrip-based DMAs. However, the meta-atoms' inherent radiative properties are arguably more naturally captured by the dipole concept.
\end{rem}

\subsection{Insights into MC}

To work out the dependence of the radiation pattern on the DMA configuration more explicitly, we use the block-matrix inversion lemma and the Woodbury identity. We define
\begin{subequations}
\begin{equation}
    \tilde{\mathbf{W}} = \mathbf{W}^0_\mathcal{MM} - \mathbf{W}^0_\mathcal{MF} \left( \mathbf{W}^0_\mathcal{FF} \right)^{-1} \mathbf{W}^0_\mathcal{FM},
    \label{eq8aaa}
\end{equation}
    \begin{equation}
    \begin{split}
        \mathbf{\Omega}_\mathcal{MM}(\mathbf{c}) &= \left[ \left(\mathbf{W} (\mathbf{c}) \right)^{-1} \right]_\mathcal{MM} =\left(  \tilde{\mathbf{W}} - \mathrm{diag}(\mathbf{c}) \right)^{-1} \\&= \Bigl[
         \tilde{\mathbf{W}}^{-1}\!+\!\tilde{\mathbf{W}}^{-1}
         \bigl([\mathrm{diag}(\mathbf{c})]^{-1}\!-\!\tilde{\mathbf{W}}^{-1}\bigr)^{-1}
         \tilde{\mathbf{W}}^{-1}
       \Bigr],
       \end{split}
        \label{eq8bbbb}
    \end{equation}
\end{subequations}
and find
\begin{equation}
    \bigl[\mathbf{W}(\mathbf{c})^{-1}\bigr]_{\mathcal{MF}}
    = -\,\mathbf{\Omega}_{\mathcal{MM}}(\mathbf{c})
       \,\mathbf{W}^0_{\mathcal{MF}}\,
       \bigl(\mathbf{W}^0_{\mathcal{FF}}\bigr)^{-1} .
\label{eq99b9}
\end{equation}
Substituting Eq.~(\ref{eq99b9}) into Eq.~(\ref{eq7}) and using the expression in the second line of Eq.~(\ref{eq8bbbb}) for $\mathbf{\Omega}_\mathcal{MM}(\mathbf{c})$, we obtain
\begin{equation} \begin{split}
    e^{\mathrm{rad},k}(\mathbf{r},\mathbf{c}) = -\left( \mathbf{v}_k(\mathbf{r})\right)^\top\! \tilde{\mathbf{W}}^{-1} \mathbf{W}^0_\mathcal{MF} \!\left(\mathbf{W}^0_\mathcal{FF}\right)^{-1}\!\mathbf{e}^\mathrm{inc}_\mathcal{F}   \\ + \left( \mathbf{v}_k(\mathbf{r})\right)^\top\! \tilde{\mathbf{W}}^{-1}\! \bigl(\!\tilde{\mathbf{W}}^{-1}
               -\!\mathrm{diag}(\mathbf{c})^{-1}\bigr)^{-1}
         \tilde{\mathbf{W}}^{-1} \mathbf{W}^0_\mathcal{MF} \!\left(\mathbf{W}^0_\mathcal{FF}\right)^{-1} \!\mathbf{e}^\mathrm{inc}_\mathcal{F}.
    \end{split}
    \label{eqq10}
\end{equation}
The first [second] term on the right hand side in Eq.~(\ref{eqq10}) represents contributions to the radiated field that are independent [dependent] of the DMA configuration $\mathbf{c}$. The dependence on $\mathbf{c}$ is seen to generally be non-linear due to a matrix inversion. By inspection of Eq.~(\ref{eq8bbbb}), we observe that the off-diagonal elements of $\tilde{\mathbf{W}}$ mediate the coupling between the different entries of $\mathbf{c}$. This mathematical observation precisely matches the physical interpretation. The off-diagonal elements of the first term on the right hand side in Eq.~(\ref{eq8aaa}) contain the background Green's functions between the meta-atoms which capture the coupling between them that is mediated via the waveguide or cavity; meanwhile, the indirect coupling between meta-atoms via the feeds is captured by the second term on the right hand side in Eq.~(\ref{eq8aaa}). 

The commonly assumed simplified DMA model with a linear mapping from configuration to radiation pattern~\cite{smith2017analysis,yurduseven2018dynamically,shlezinger2019dynamic,wang2020dynamic,shlezinger2021dynamic,boyarsky2021electronically,RISnDMA_2024,DMA2023,jabbar202460,DMA_WPT} arises from assumptions of negligible MC and can be recovered from our physics-consistent 
model in the limiting case in which the off-diagonal entries of $\tilde{\mathbf{W}}$ vanish. The first line of Eq.~(\ref{eq8bbbb}) reveals that
\begin{equation}
    \lim\limits_{ \tilde{\mathbf{W}} \to {\mathrm{diag}(\mathbf{d})} } \!\!\!\!\!\!
    \left[ \mathbf{\Omega}_\mathcal{MM}(\mathbf{c}) \right] = -\left[  \mathrm{diag}(\mathbf{c}-\mathbf{d})\right]^{-1} =  -\mathrm{diag}(\mathbf{c}^\prime),
\end{equation}
where $\mathbf{d}\in\mathbb{C}^{N_\mathrm{M}}$ is a vector containing the diagonal entries of $\tilde{\mathbf{W}}$, and the $i$th entry of $\mathbf{c}^\prime$ is the inverse of the difference between the $i$th entry of $\mathbf{c}$ and the $i$th entry of $\mathbf{d}$. Hence, in the limit of vanishing MC we recover the common simplified DMA model:   
\begin{equation}
    \lim\limits_{\tilde{\mathbf{W}} \to {\mathrm{diag}(\mathbf{d})}} \!\!\!
    \left[ e^{\mathrm{rad},k}(\mathbf{r},\mathbf{c}) \right] 
    \!= \!\left( \mathbf{v}_k(\mathbf{r}) \right)^\top\! \!
    \mathrm{diag}(\mathbf{c}^\prime) \mathbf{W}^0_\mathcal{MF}\! \left(\mathbf{W}^0_\mathcal{FF}\right)^{-1}\!
    \!\mathbf{e}^\mathrm{inc}_\mathcal{F}.
    \label{eq_simplifiedModel}
\end{equation}
In applications of Eq.~(\ref{eq_simplifiedModel}) to stacked-microstrip DMAs, the entries of $\mathbf{W}^0_\mathcal{MF}$ are usually identified based on simple 1D waveguide transmission properties, and $\left(\mathbf{W}^0_\mathcal{FF}\right)^{-1}$ is approximated as an identity matrix because each feed is on a separate microstrip (see Fig.~\ref{Fig1}) and the microstrips are assumed not to couple to each other.  
An alternative route to recover the common simplified model from our physics-consistent model consists in imposing simultaneously that $G_{ij} = 0$ for $i,j\in\mathcal M,\ i\neq j$ \textit{and} $G_{ij}=0$ for $i\in\mathcal F,\ j\in\mathcal M$ (while $G_{ji}\neq 0$ for $j\in\mathcal M,\ i\in\mathcal F$). This alternative route resembles the ``unilateral approximation''~\cite{ivrlavc2010toward} that prevents coupling between the meta-atoms via bounces off the feeds. (Note that this approximation introduces a non-reciprocity.) Clearly, by construction these simplified models cannot capture potential benefits of MC in optimizing DMAs for wireless communications.

To systematically investigate the latter, a quantification of the DMA's MC strength is required. Based on our system model insights, we quantify the DMA's \textit{absolute} MC strength with
\begin{equation}
\mu = \left\lVert \tilde{\mathbf{W}} - \mathrm{diag}\left( \tilde{\mathbf{W}} \right) \right\rVert_2.
\end{equation}
The ease of inverting $\tilde{\mathbf{W}}-\mathrm{diag}(\mathbf{c})$, and hence the strength of the non-linearity in the mapping from $\mathbf{c}$ to $\mathbf{\Omega}_\mathcal{MM}(\mathbf{c})$ and ultimately $e^{\mathrm{rad},k}(\mathbf{r},\mathbf{c})$, depends on the DMA's \textit{relative} MC strength: 
\begin{equation}
    \mu_\mathrm{n} = \frac{\mu}{ \left\lVert \mathrm{diag}\left( \tilde{\mathbf{W}} - \mathrm{diag}(\mathbf{c}) \right) \right\rVert}_2.
\end{equation}

\subsection{Summary}
In this section, we have introduced a compact and implementation-agnostic DMA system model that is consistent with physics. We investigated how the model's mathematical structure results in intertwined influences of the meta-atoms on the radiation pattern. We also clarified how to recover common simplified models from our physics-consistent model by neglecting MC. Finally, we introduced a metric to quantify the MC strength.

\section{Theory}
\label{sec_Theory}

Equipped with the system model from Sec.~\ref{sec_system_model}, in this section we theoretically analyze the \textit{strength} and \textit{simplicity} of the dependence of the DMA radiation pattern $e^{\mathrm{rad},k}(\mathbf{r},\mathbf{c})$ on the DMA configuration $\mathbf{c}$. To that end, we analytically derive the first and second partial derivatives of $e^{\mathrm{rad},k}(\mathbf{r},\mathbf{c})$ with respect to $\mathbf{c}$. The first derivative (Jacobian) informs us about the sensitivity of $e^{\mathrm{rad},k}(\mathbf{r},\mathbf{c})$ to $\mathbf{c}$. The larger the Jacobian is, the \textit{stronger} $e^{\mathrm{rad},k}(\mathbf{r},\mathbf{c})$ depends on $\mathbf{c}$. We hypothesize that a larger sensitivity translates into a larger ability to tailor $e^{\mathrm{rad},k}(\mathbf{r},\mathbf{c})$ to a desired wireless functionality; we confirm this hypothesis numerically and experimentally in the subsequent sections. The second derivative (Hessian) informs us about how the dependence of $e^{\mathrm{rad},k}(\mathbf{r},\mathbf{c})$ on a given entry of $\mathbf{c}$ depends itself on another given entry of $\mathbf{c}$. The Hessian is the first term in the Taylor expansion of $e^{\mathrm{rad},k}(\mathbf{r},\mathbf{c})$ with respect to $\mathbf{c}$ that indicates curvature. Thus, the larger the Hessian is, the \textit{less simple} (i.e., more non-linear) the mapping from $\mathbf{c}$ to $e^{\mathrm{rad},k}(\mathbf{r},\mathbf{c})$ is. 

For simplicity, we consider from now on a DMA with a single feed (i.e., $N_\mathrm{F}=1$). The analysis of the multiple-feed case is deferred to future work due to the complexity of considering the influence of the precoder on the partial derivative of the radiation pattern with respect to the DMA configuration. We anticipate that the key physical insights presented here will carry over to the multi-feed case. 
For $N_\mathrm{F}=1$, $\mathbf{e}^\mathrm{inc}_\mathcal{F}$ is a scalar that we denote by $e^\mathrm{inc}$.

\subsection{Sensitivity of  Radiation Pattern to  Configuration}
\label{subsec_II_sensitivity}

To obtain the partial derivative of $e^{\mathrm{rad},k}(\mathbf{r},\mathbf{c})$ with respect to $\mathbf{c}$ from Eq.~(\ref{eq7}), we first note that
\begin{equation}
     \mathbf{j}^{\mathrm{rad},k}(\mathbf{r},\mathbf{c}) = \frac{\partial e^{\mathrm{rad},k} (\mathbf{r},\mathbf{c})}{\partial \mathbf{c}}  = \mathbf{v}_k^\top (\mathbf{r}) \mathbf{J}^\mathrm{p}(\mathbf{c})  \in \mathbb{C}^{N_\mathrm{M}} ,
     \label{eq900}
\end{equation}
where $\mathbf{J}^\mathrm{p}(\mathbf{c}) \in \mathbb{C}^{N_\mathrm{M} \times N_\mathrm{M}}$ is the Jacobian of the meta-atoms' dipole moment, i.e., the $i$th column of $\mathbf{J}^\mathrm{p}(\mathbf{c})$ is $\mathbf{j}^\mathrm{p}_i(\mathbf{c}) = \frac{\partial}{\partial c_i} \left(  \left[\left(\mathbf{W}(\mathbf{c})\right)^{-1} \right]_\mathcal{MF} \right)e^\mathrm{inc} \in \mathbb{C}^{N_\mathrm{M} }$ (recall that we chose $N_\mathrm{F}=1$). For notational ease, we define $\mathbf{\Omega}(\mathbf{c}) =  \left(\mathbf{W}(\mathbf{c})\right)^{-1}$. After a few linear algebra steps detailed in Appendix~\ref{AppendixA}, we obtain
\begin{subequations}\label{eq1000}
   \begin{equation}
       \mathbf{j}^\mathrm{p}_i(\mathbf{c}) = \mathbf{\Omega}_{\mathcal{M}\mathcal{M}_i}(\mathbf{c}) \mathbf{p}_{\mathcal{M}_i}(\mathbf{c}) , \tag{\theequation a}
       \label{eq1000a}
   \end{equation} 
   \begin{equation}
       \mathbf{J}^\mathrm{p}(\mathbf{c}) = \mathbf{\Omega}_\mathcal{MM}(\mathbf{c}) \mathrm{diag}(\mathbf{p}_\mathcal{M}(\mathbf{c})) . \tag{\theequation b}
       \label{eq1000b}
   \end{equation}
\end{subequations}
For clarity, we recall that  in Eq.~(\ref{eq1000}) the cardinality of the sets $\mathcal{M}_i$ and $\mathcal{F}$ is unity, and that $\mathbf{p}_\mathcal{M}(\mathbf{c}) = \mathbf{\Omega}_\mathcal{MF}(\mathbf{c}) e^\mathrm{inc}$. 
Substituting Eq.~(\ref{eq1000b}) into Eq.~(\ref{eq900}), we have
\begin{equation}
     \mathbf{j}^{\mathrm{rad},k}(\mathbf{r},\mathbf{c})  =  \mathbf{v}_k^\top (\mathbf{r}) \mathbf{\Omega}_{\mathcal{M}\mathcal{M}}(\mathbf{c}) \mathrm{diag}(\mathbf{p}_\mathcal{M}(\mathbf{c}))  ,
     \label{eq1100}
\end{equation}
where we observe that the partial derivative of the radiation pattern with respect to $\mathbf{c}$ is in fact a function of $\mathbf{c}$. 

The \textit{absolute} radiation pattern sensitivity can hence be quantified by $\|\mathbf{j}^{\mathrm{rad},k}(\mathbf{r},\mathbf{c})\|_2$ based on Eq.~(\ref{eq1100}). However, at this stage, some care is required to ensure a fair comparison between DMAs with different MC strengths because they likely differ regarding their radiation efficiency. Our sensitivity metric should be agnostic to the DMA's radiation efficiency so that it informs us about the \textit{relative} (as opposed to \textit{absolute}) change of the radiation pattern in response to a change in DMA configuration. Thus, a normalization of $\|\mathbf{j}^{\mathrm{rad},k}(\mathbf{r},\mathbf{c})\|_2$ is necessary.

The only analytically tractable normalization we are aware of consists in independently normalizing each column of $\mathbf{J}^\mathrm{p}(\mathbf{c})$ via division by the corresponding entry of $\mathbf{p}_\mathcal{M}(\mathbf{c})$, yielding
\begin{equation}
    \hat{\mathbf{J}}^\mathrm{p}(\mathbf{c}) = \mathbf{J}^\mathrm{p}(\mathbf{c}) \left[ \mathrm{diag}(\mathbf{p}_\mathcal{M}(\mathbf{c})) \right]^{-1} = \mathbf{\Omega}_\mathcal{MM}(\mathbf{c}) .
    \label{eqJPP}
\end{equation}
Thus, $\|\hat{\mathbf{J}}^\mathrm{p}(\mathbf{c})\|_2 = \|\mathbf{\Omega}_\mathcal{MM}(\mathbf{c})\|_2$. In Appendix~\ref{AppendixB}, we derive an upper bound for $\|\mathbf{\Omega}_\mathcal{MM}(\mathbf{c})\|_2$:
\begin{equation}
    \|\mathbf{\Omega}_\mathcal{MM}(\mathbf{c})\|_2 \leq  \frac{1}{\sigma_{N_\mathrm{M}}(1-\tilde{\mu})},
    \label{eq_bound}
\end{equation}
where $\tilde{\mu} = \mu_\mathrm{n} \frac{\sigma_1}{\sigma_{N_\mathrm{M}}}$ and $\sigma_i$ is the $i$th singular value (in descending order) of $\mathrm{diag}( \tilde{\mathbf{W}} - \mathrm{diag}(\mathbf{c}) )$.
Note that $\frac{\sigma_1}{\sigma_{N_\mathrm{M}}}$ is the condition number of $\mathrm{diag}( \tilde{\mathbf{W}} - \mathrm{diag}(\mathbf{c}) )$ which is simply the ratio of the modulus of the largest and smallest diagonal entries in this case.
Applying the submultiplicativity property to $\hat{\mathbf{j}}^{\mathrm{rad},k}(\mathbf{r},\mathbf{c}) = \mathbf{v}_k^\top (\mathbf{r}) \hat{\mathbf{J}}^\mathrm{p}(\mathbf{c}) e^\mathrm{inc}$, it follows that
\begin{equation}
     \|\hat{\mathbf{j}}^{\mathrm{rad},k}(\mathbf{r},\mathbf{c})\|_2 \leq \frac{\|\mathbf{v}_k\|_2 |e^\mathrm{inc}|}{ \sigma_{N_\mathrm{M}}(1-\tilde{\mu})} .
\end{equation}
By definition, $\tilde{\mu}>0$; moreover, according to our numerical results in the next section, $\tilde{\mu}$ remains below unity in typical DMA architectures that we consider. For $0<\tilde{\mu}<1$, the larger $\tilde{\mu}$ is, the higher is the upper bound on $\|\mathbf{\Omega}_\mathcal{MM}(\mathbf{c})\|_2$ according to Eq.~(\ref{eq_bound}) and hence on $\|\hat{\mathbf{J}}^\mathrm{p}(\mathbf{c})\|_2$ according to Eq.~(\ref{eqJPP}). In other words, the upper bound on the sensitivity increases with the relative MC strength. We further note that the upper bound is inversely proportional to $\sigma_{N_\mathrm{M}}$, and thus it increases with the largest ``dressed'' polarizability (which combines the effect of the polarizability and the self-interactions originating from $G_{ii}$~\cite{yoo2012effective}) of the meta-atoms. 

As explained in Appendix~\ref{AppendixB}, the bound could be attained with equality if the coupling between the meta-atoms was controllable. One possible route toward such controllable couplings would be the realization of a ``beyond-diagonal DMA'' (similar to BD-RIS~\cite{del2024physics,nerini2024global}) equipped with additional  tunable lumped elements inside the cavity that do not radiate to the scene. An initial exploration of BD-DMAs can be found in our recent follow-up work~\cite{prod2025beyond}. 

Altogether, the theoretical analysis in this subsection suggests that the dependence of the DMA radiation pattern on the DMA configuration gets stronger as MC in the DMA gets stronger, even after normalization to remove the effect of the DMA's radiation efficiency. We confirm this relation through extensive numerical and experimental studies in the next sections.

\subsection{Non-Linearity of Radiation Pattern Parametrization}

The Hessian $\mathbf{H}^\mathrm{p} = \frac{\partial \mathbf{J}^\mathrm{p}}{\partial \mathbf{c}}\in \mathbb{C}^{N_\mathrm{M} \times N_\mathrm{M} \times N_\mathrm{M}}$ of the meta-atoms' dipole moments is a rank-three tensor. We begin by evaluating the $i$th component of the normalized Hessian:
\begin{equation}
    \hat{\mathbf{H}}_i^\mathrm{p}(\mathbf{c}) = \frac{\partial \hat{\mathbf{J}}^\mathrm{p}(\mathbf{c})}{\partial c_i} = \mathbf{\Omega}_\mathcal{MM}(\mathbf{c}) \mathbf{E}_{ii}  \mathbf{\Omega}_\mathcal{MM}(\mathbf{c}) \in \mathbb{C}^{N_\mathrm{M}\times N_\mathrm{M}},
    \label{Hpp}
\end{equation}
where $\mathbf{E}_{ii}$ denotes an $N_\mathrm{M} \times N_\mathrm{M}$ matrix whose entries are zero except for the $i$th diagonal entry which is unity. Applying submultiplicativity arguments to Eq.~(\ref{Hpp}) yields $\|\hat{\mathbf{H}}_i^\mathrm{p}(\mathbf{c})\|_2 \leq \|\mathbf{\Omega}_\mathcal{MM}(\mathbf{c})\|_2^2$ which we combine with Eq.~(\ref{eq_bound}) to obtain:
\begin{equation}
    \|\hat{\mathbf{H}}_i^\mathrm{p}(\mathbf{c})\|_2 \leq \frac{1}{\sigma_{N_\mathrm{M}}^2(1-\tilde{\mu})^2}.
\end{equation}
It further follows from applying the submultiplicativity property to $\hat{\mathbf{H}}_i^{\mathrm{rad},k}(\mathbf{r},\mathbf{c}) = \mathbf{v}_k^\top (\mathbf{r}) \hat{\mathbf{H}}_i^\mathrm{p}(\mathbf{c}) e^\mathrm{inc}$ that
\begin{equation}
     \|\hat{\mathbf{H}}_i^{\mathrm{rad},k}(\mathbf{r},\mathbf{c})\|_2 \leq \frac{\|\mathbf{v}_k\|_2 |e^\mathrm{inc}|}{ \sigma_{N_\mathrm{M}}^2(1-\tilde{\mu})^2} .
\end{equation}
As the MC strength $\tilde{\mu}$ increases, the upper bound on $\|\hat{\mathbf{H}}_i^{\mathrm{rad},k}(\mathbf{r},\mathbf{c})\|_2$ increases even faster than the upper bound on $\|\hat{\mathbf{j}}^{\mathrm{rad},k}(\mathbf{r},\mathbf{c})\|_2$. Our analysis hence suggests that the non-linearity in the mapping from $\mathbf{c}$ to $e^{\mathrm{rad},k}(\mathbf{r},\mathbf{c})$ gets stronger as the MC strength is increased.

\subsection{Summary}
We showed in this section that the upper bounds on the normalized Jacobian and Hessian of the DMA radiation pattern with respect to the DMA configuration increase with the DMA's relative MC strength. These analytical findings suggest a \textit{fundamental trade-off between strength and simplicity} of the dependence of the DMA radiation pattern on the DMA configuration. Earlier works~\cite{del2021deeply,prod2024mutual,nerini2024global,semmler2024decoupling}, as well as~\cite{liu2025optimization} whose preprint appeared shortly after the preprint of the present paper, already provided related observations regarding this trade-off in various contexts. In the next sections, we provide exhaustive systematic numerical and experimental evidence for this trade-off in the DMA context. Intuitively, the trade-off is expected based on a multi-bounce picture~\cite{rabault2024} of wave propagation in the DMA: In a DMA with stronger MC, multi-bounce wave trajectories are more prominent; the more often a wave trajectory bounces off meta-atoms, the more it is affected by their configuration and the more intertwined the effects of the meta-atoms' configuration on the trajectory are~\cite{prod2024mutual}.

\section{Full-Wave Numerical Validation}
\label{sec_numerical}

In this section, we numerically demonstrate in full-wave simulations that an increase in the DMA MC strength results in both a stronger DMA radiation pattern sensitivity and a stronger non-linearity of the DMA radiation pattern parametrization, as well as a greater ability to tailor the DMA radiation pattern to the needs of a wireless functionality. We evaluate sensitivity and non-linearity in terms of the model-based metrics from our theoretical analysis in Sec.~\ref{sec_Theory} as well as in terms of model-free metrics that we use for our experimental validation in Sec.~\ref{sec_experimental}, confirming that both types of metrics show the same trends as a function of MC strength.
Our numerical study is not intended to reproduce every detail of our subsequent experiment.

\begin{figure}
    \centering
    \includegraphics[width=\columnwidth]{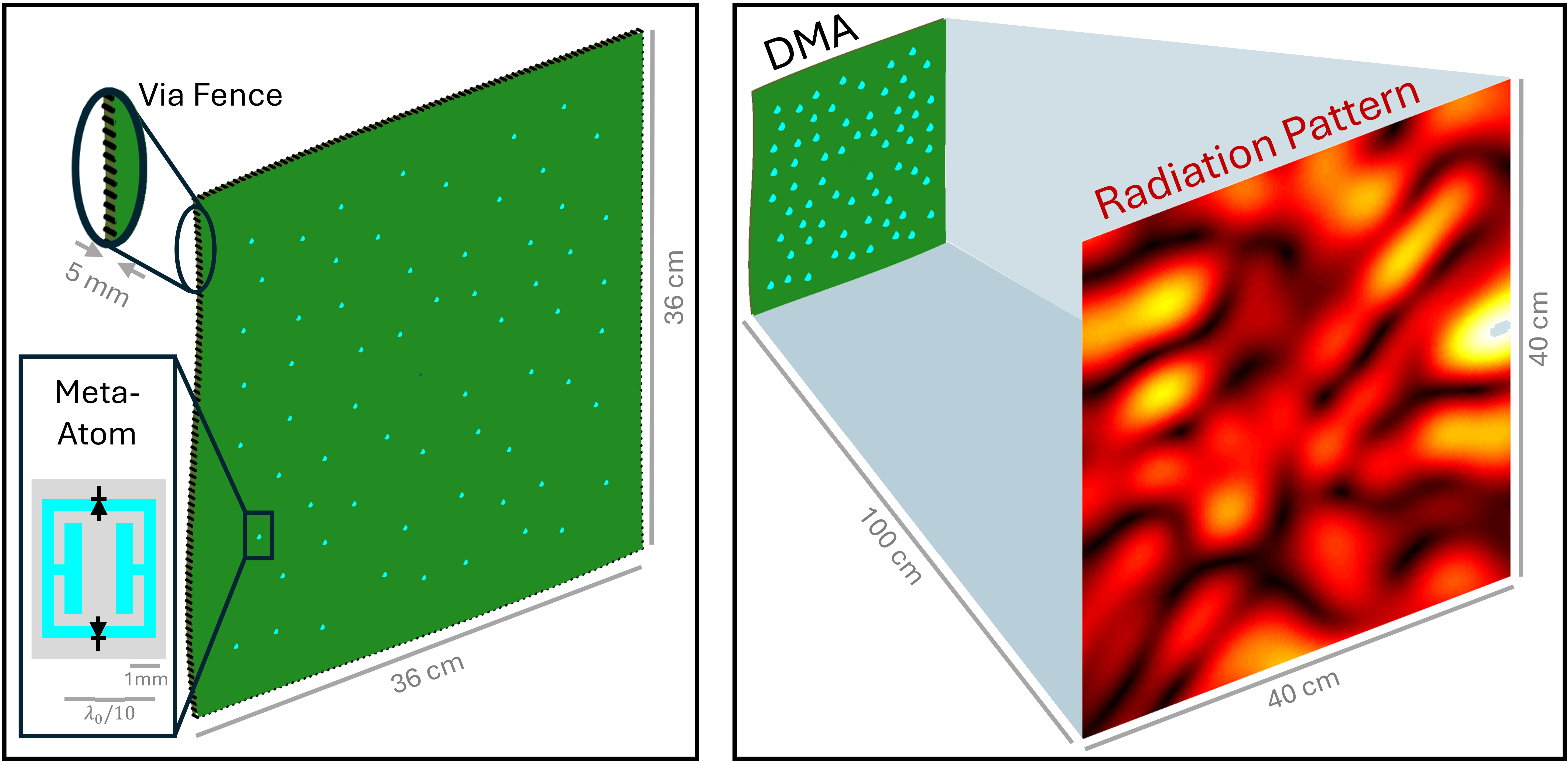}
    \caption{DMA design and setup in the numerical full-wave simulations.}
    \label{FigY}
\end{figure}

\subsection{Full-Wave Simulation Details}

We consider the chaotic-cavity-backed DMA displayed in Fig.~\ref{FigY}, similar to the one proposed in~\cite{sleasman2020implementation}.
The DMA consists of a quasi-2D chaotic cavity whose boundaries are defined by two parallel conducting layers separated by a distance of 5~mm, and an irregularly shaped fence of 200 vias between the two conducting layers. Because of the via fence's irregular shape (and the presence of the scattering meta-atoms), wave chaos arises in the cavity. The DMA's cavity is excited at its center by a single feed from the DMA's back. 64 meta-atoms are randomly placed on the DMA's front surface. The meta-atoms are typical complementary electric-LC (cELC) resonators (see inset in Fig.~\ref{FigY} and~\cite{pulido2017polarizability,del2020learned}), parametrized by PIN diodes so that they are 1-bit programmable. In its ``ON'' state, a meta-atom significantly leaks out energy from the cavity to the scene. 

We use an in-house developed full-wave solver based on the coupled-dipole formalism.\footnote{The coupled-dipole formalism is a full-wave solver because it rigorously solves Maxwell’s equations, self-consistently capturing all near- and far-field interactions among the dipoles, fully accounting for scattering, radiation, and interference effects.} Similar to~\cite{yoo2019analytic}, we model feed, vias and meta-atoms as dipoles that are coupled by the 2D Green's functions relevant to the parallel-plate waveguide geometry. Using the reduced-basis representation introduced in~\cite{prod2023efficient}, we deduce the \textit{background} Green's functions to be inserted into our DMA system model from Sec.~\ref{sec_system_model}. The meta-atom's polarizability was extracted based on finite-element simulations in~\cite{pulido2017polarizability}. Our full-wave solver is implemented in an automatic differentiation framework (JAX, Google Research) that gives us direct access to gradients. 

To consider different levels of MC strength in the DMA, we alter the lossiness of the substrate in the DMA's cavity. We use $k_0(1 - \jmath\gamma)$ as the substrate wavenumber, where $k_0$ is the wavenumber in the absence of loss, and $\gamma$ controls the level of lossiness. We found that the dependence of the MC strength on $\gamma$ is smoother than on the number of vias in the fence (the latter was used in~\cite{prod2024mutual}).

\subsection{Optimization}

To assess the influence of the DMA's MC strength on the ability to tailor its radiation pattern to the requirements of a wireless functionality, we consider the prototypical problem of enhancing the gain of a single-user channel. 
This metric is a commonly considered key performance indicator because it directly correlates with end-to-end communication metrics (e.g., capacity, bit error rate, or throughput). We emphasize that our goal of maximizing the channel gain is \textit{not} equivalent to other conceivable goals such as maximizing directivity or minimizing beam width that are left for future work. In addition, explorations of multi-user scenarios are deferred to future work because of their complexity and specificity to choices like the noise level.
The DMA is deployed as base station, and the user equipment (UE) is at a fixed location $\mathbf{r}_\mathrm{UE}$ in the scene. The channel gain is then directly proportional to the intensity of the component of the DMA's radiation pattern at the UE location and parallel to the UE antenna's orientation: $|h(\mathbf{r}_\mathrm{UE},k,\mathbf{c})|^2 \propto |e^{\mathrm{rad},k}(\mathbf{r}_\mathrm{UE},\mathbf{c})|^2$, where $h(\mathbf{r}_\mathrm{UE},k,\mathbf{c})\in\mathbb{C}$ is the single-input single-output wireless channel from the DMA's port to the UE antenna's port.

To solve the non-linear, non-convex and high-dimensional optimization problem of maximizing channel gain subject to binary constraints on the DMA configuration, we use the coordinate ascent method summarized in Algorithm~\ref{Alg_CoordAsc}. The objective function that we seek to maximize is $\mathcal{C}=|e^{\mathrm{rad},k}(\mathbf{r}_\mathrm{UE},\mathbf{c})|^2$.

\begin{algorithm}
\footnotesize
\label{Alg_CoordAsc}
Evaluate the objective functions $\left\{\mathcal{C}\right\}_{\mathrm{init}}$ for $512$ random DMA configurations $\left\{\mathbf{c}\right\}_{\mathrm{init}}$.\\
Select $\mathcal{C}_{\rm{curr}}$ as the maximum from $\left\{\mathcal{C}\right\}_{\mathrm{init}}$ and $\mathbf{c}_{\mathrm{curr}}$ from $\left\{\mathbf{c}\right\}_{\mathrm{init}}$ at the same index such that $\mathcal{C}_{\mathrm{curr}} = \mathcal{C}\left(\mathbf{c}_{\mathrm{curr}}\right) = \mathrm{max}\left(\left\{\mathcal{C}\right\}_{\mathrm{init}}\right)$.\\
$\ell \gets 0$, $i \gets 0$\\
\While{$\ell < N_\mathrm{M}$}{
    $i \gets i+1$\\
    $\mathbf{c}_\mathrm{temp} \gets \mathbf{c}_\mathrm{curr}$  with the $(\mathrm{mod}(i-1,N_{\rm M})+1)$th bit flipped.\\
    $\mathcal{C}_{\rm temp} \gets \mathcal{C}\left(\mathbf{c}_\mathrm{temp}\right)$.\\
    \eIf{
    $\mathcal{C}_{\rm temp} > \mathcal{C}_{\rm curr}$
    } {
    $\mathbf{c}_\mathrm{curr} \gets \mathbf{c}_\mathrm{temp}$ \\
    $\mathcal{C}_\mathrm{curr} \gets \mathcal{C}_\mathrm{temp}$ \\
    $\ell \gets 0$
    } {$\ell \gets \ell+1$}
    }
\KwOut{$\mathbf{c}_{\rm curr}$ and $\mathcal{C}_\mathrm{curr}$.}
\caption{Channel Gain Enhancement by Coordinate Ascent.}
\end{algorithm}

\subsection{Numerically Evaluated Metrics}
\label{sec_metrics_numerics}

For each considered level of substrate lossiness, we evaluate the following metrics.

\subsubsection{MC Strength} 

Having direct access to $\tilde{\mathbf{W}}$ and $\mathbf{c}$, we directly evaluate $\mu_\mathrm{n}(\mathbf{c})$ and $\tilde{\mu}(\mathbf{c})$ for 512 random realizations of $\mathbf{c}$, and we denote by $\langle\mu_\mathrm{n}\rangle$ and $\langle\tilde{\mu}\rangle$ the respective averages over the realizations of $\mathbf{c}$. 

\subsubsection{Radiation Pattern Sensitivity} Our solver's implementation in an automatic differentiation framework gives us direct access to the field $e^{\mathrm{rad},k}(\mathbf{r},\mathbf{c})$ and its partial derivative $\mathbf{j}^{\mathrm{rad},k}(\mathbf{r},\mathbf{c})$ at $48 \times 48$ regularly spaced grid positions $\mathbf{r}$ spanning a $0.4\times 0.4\ \mathrm{m}^2$ square area located 1~m in front of the DMA (see right panel in Fig.~\ref{FigY}), i.e., in the radiating near field (DMA aperture dimension: 30~cm; DMA operation frequency: 10~GHz). In addition, since we have access to the dipole moments, we can evaluate $\hat{\mathbf{j}}^{\mathrm{rad},k}(\mathbf{r},\mathbf{c})$. We compute these three quantities for 512 random realizations of $\mathbf{c}$. Then, we evaluate two sensitivity metrics:
\begin{subequations}    \label{eq21}
    \begin{equation}
        \langle\hat{\sigma}\rangle = \langle\|\hat{\mathbf{j}}^{\mathrm{rad},k}(\mathbf{r},\mathbf{c})\|_2\rangle,
        \label{eq21a}
    \end{equation}
    \begin{equation}
        \langle\sigma\rangle = \frac{\langle | \mathbf{j}^{\mathrm{rad},k}_i(\mathbf{r},\mathbf{c}) | \rangle}{\langle | e^{\mathrm{rad},k}(\mathbf{r},\mathbf{c}) | \rangle},
        \label{eq21b}
    \end{equation}
\end{subequations}
\noindent where the average on the right hand side is over 512 random realizations of $\mathbf{c}$ and the $48\times48$ considered positions $\mathbf{r}$ in Eq.~(\ref{eq21a}), and additionally over the entries of $\mathbf{j}^{\mathrm{rad},k}(\mathbf{r},\mathbf{c})$ in Eq.~(\ref{eq21b}). $\hat{\sigma}$ is the metric used in our theoretical analysis in Sec.~\ref{sec_Theory} and can only be evaluated given the system model; meanwhile, $ \langle\sigma\rangle$ is a model-agnostic metric whose finite-difference approximation we also evaluate experimentally in Sec.~\ref{sec_experimental}.

\subsubsection{Linearity of Radiation Pattern Parametrization} Analogous to Eq.~(\ref{eq21}), we evaluate
\begin{subequations} \label{eq22}
    \begin{equation}
        \langle\hat{\theta}\rangle = \langle\|\hat{\mathbf{H}}^{\mathrm{rad},k}(\mathbf{r},\mathbf{c})\|_2\rangle,
        \label{eq22a}
    \end{equation}
    \begin{equation}
        \langle\theta\rangle = \frac{\langle | \mathbf{H}^{\mathrm{rad},k}_{i,j}(\mathbf{r},\mathbf{c}) | \rangle}{\langle | e^{\mathrm{rad},k} (\mathbf{r},\mathbf{c})| \rangle}.
        \label{eq22b}
    \end{equation}
\end{subequations}
In Eq.~(\ref{eq22b}), we only average over the off-diagonal entries of $\mathbf{H}^{\mathrm{rad},k}(\mathbf{r},\mathbf{c})$, in line with the finite-difference approximation thereof that we experimentally evaluate in Sec.~\ref{sec_experimental}. 
In addition, we evaluate the linearity metric
\begin{equation}
    \langle \zeta \rangle = \left\langle \frac{\text{SD}\left[e^{\mathrm{rad},k}(\mathbf{r},\mathbf{c})\right]}{\text{SD}\left[e^{\mathrm{rad},k}(\mathbf{r},\mathbf{c})-\epsilon^{\mathrm{rad},k}(\mathbf{r},\mathbf{c})\right]} \right\rangle
\end{equation}
where the standard deviation is taken over 512 random DMA configurations $\mathbf{c}$ and the mean is taken over the $48\times48$ considered positions $\mathbf{r}$. $\epsilon^{\mathrm{rad},k}(\mathbf{r},\mathbf{c})$ denotes the prediction of $e^{\mathrm{rad},k}(\mathbf{r},\mathbf{c})$ based on a linear regression. $\langle \zeta \rangle$ is insensitive to radiation that does not depend on $\mathbf{c}$. The larger $\langle \zeta \rangle$ is, the more linear (or affine) the mapping from $\mathbf{c}$ to $e^{\mathrm{rad},k}(\mathbf{r},\mathbf{c})$ is. Further discussion of this metric is available in Sec. II of~\cite{rabault2024}. 

\subsubsection{Channel Gain Enhancement} We quantify the achieved channel gain enhancement as 
\begin{equation}
    \langle\eta\rangle = \frac{\langle |e^{\mathrm{rad},k}(\mathbf{r},\mathbf{c}_\mathrm{opt})|^2 \rangle}{\langle |e^{\mathrm{rad},k}(\mathbf{r},\mathbf{c}_\mathrm{rand})|^2 \rangle}
\end{equation}
where $\mathbf{c}_\mathrm{opt}$ is the optimized DMA configuration identified with Algorithm~\ref{Alg_CoordAsc}, and the averages on the right hand side are over the $48\times48$ considered positions $\mathbf{r}$ for the numerator and additionally over 512 random configurations $\mathbf{c}_\mathrm{rand}$ for the denominator.
Provided the transmit power $P_\mathrm{T}$ and receiver noise power $P_\mathrm{N}$ are fixed, $\langle\eta\rangle$ equals the average signal-to-noise ratio (SNR) enhancement enabled by optimizing the DMA configuration because the SNR is proportional to the channel gain: $\mathrm{SNR} = P_\mathrm{T} |e^{\mathrm{rad},k}(\mathbf{r},\mathbf{c})|^2  /P_\mathrm{N}$.

\subsection{Numerical Full-Wave Results}

\begin{figure}
    \centering
    \includegraphics[width=0.5\columnwidth]{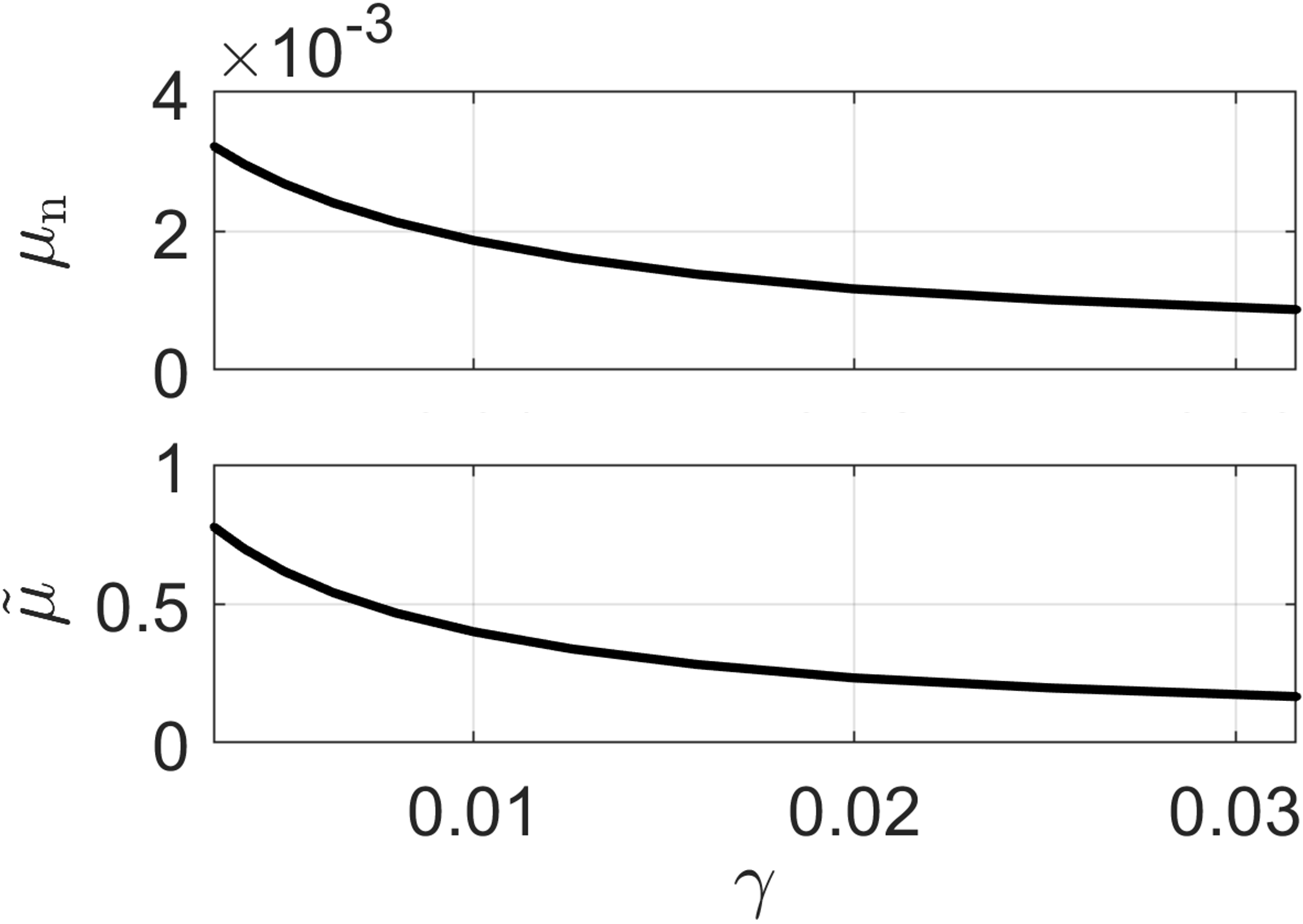}
    \caption{DMA MC strength vs substrate lossiness factor $\gamma$.}
    \label{FigZZ}
\end{figure}

We first inspect in Fig.~\ref{FigZZ} the influence of the substrate lossiness factor $\gamma$ on the DMA's MC strength $\mu_\mathrm{n}$, as well as its variant $\tilde{\mu}$. The order of magnitude of the considered values of $\gamma$ corresponds to the losses in typical printed-circuit board (PCB) substrates. As expected, we observe a monotonic decrease of the DMA's MC strength as the substrate lossiness increases. Within the explored range of $\gamma$, $\mu_\mathrm{n}$ drops by a factor of roughly four and $\tilde{\mu}<1$ such that the bounds derived in Sec.~\ref{sec_Theory} are meaningful.

\begin{figure}[h]
    \centering
    \includegraphics[width=\columnwidth]{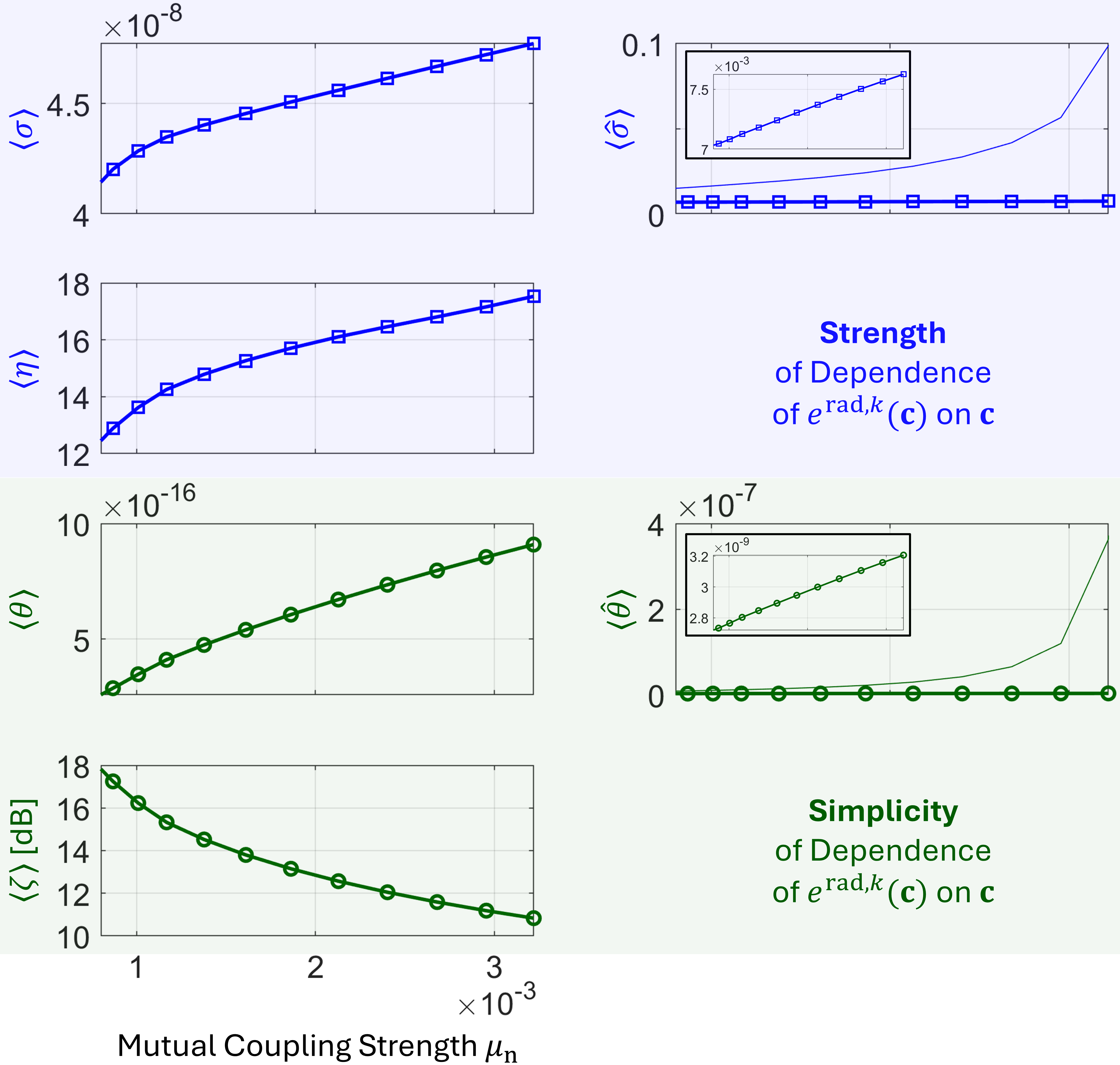}
    \caption{Numerical full-wave results for the trade-off between strength and simplicity of the dependence of $e^{\mathrm{rad},k}(\mathbf{c})$ on $\mathbf{c}$, governed by the DMA's MC strength $\mu_\mathrm{n}$. We vary $\mu_\mathrm{n}$ by altering the substrate lossiness $\gamma$ (see Fig.~\ref{FigZZ}). The left column displays (from top to bottom) the average sensitivity $\langle\sigma\rangle$, the average channel gain enhancement $\langle\eta\rangle$, the average normalized off-diagonal Hessian entries $\langle\theta\rangle$, and the average linearity metric $\langle\zeta\rangle$, as a function of MC strength. The right column displays the differently normalized average sensitivity $\langle\hat{\sigma}\rangle$ and Hessian entries  $\langle\hat{\theta}\rangle$ together with corresponding upper bounds. Insets show zoomed-in versions.} 
    \label{FigBCD}
\end{figure}

In the top row of Fig.~\ref{FigBCD}, we observe a positive correlation between the radiation pattern sensitivity and the MC strength. This general trend is clear in terms of both the average relative magnitude of the Jacobian (left), $\langle\sigma\rangle$, and the analytically tractable normalized sensitivity metric $\langle\hat{\sigma}\rangle$ (right). Our results confirm the validity of the bound on $\langle\hat{\sigma}\rangle$, and we observe the expected scaling of the bound on $\langle\hat{\sigma}\rangle$ with $\mu_\mathrm{n}$. As detailed in Appendix~\ref{AppendixB}, the bound is not tight because we cannot control the MC between the meta-atoms; as already mentioned, a beyond-diagonal DMA~\cite{prod2025beyond} can overcome this limitation and potentially attain the bound. In the second row of Fig.~\ref{FigBCD}, we see that the enhanced sensitivity under stronger MC translates into concrete benefits in wireless communications. Specifically, the achieved channel gain enhancement, $\langle\eta\rangle$, displays a clear positive correlation with the sensitivity and hence with the MC strength. We observe in Fig.~\ref{FigBCD} that roughly tripling the MC strength increases the normalized sensitivity $\langle\sigma\rangle$ by 13.6~\% which translates into a 36.0~\% increase in channel gain enhancement. In other words, stronger MC brings about substantial improvements in our ability to tailor the DMA radiation pattern to the needs of a desired wireless functionality.

The flipside of this benefit of strong MC is seen in the lower half of Fig.~\ref{FigBCD}. Stronger MC increases the non-linearity of the mapping from DMA configuration to DMA radiation pattern, as clearly revealed by all three of our monitored metrics. The average relative off-diagonal magnitudes of the Hessian (left), $\langle\theta\rangle$, as well as the analytically tractable version $\langle\hat{\theta}\rangle$ clearly increase as the MC gets stronger. Again, the validity of our bound and its scaling with $\mu_\mathrm{n}$ are confirmed. In direct correspondence with the larger normalized Hessians for larger $\mu_\mathrm{n}$, we observe a clear drop in our linearity metric $\langle\zeta\rangle$ from 17.2~dB to 10.8~dB.

\subsection{Summary}
To summarize this section, the trends in terms of how a DMA's MC strength governs the trade-off between the strength and simplicity of the dependence of the DMA radiation pattern on the DMA configuration are clearly observed, irrespective of the chosen metric. The direct translation of an improved sensitivity to an improved ability to tailor the radiation pattern to the needs of a wireless functionality is also clear. Our goal in this section was not  to reproduce our subsequent experiment in every detail, but to observe the described trends with both the theoretically and experimentally convenient metrics, as we did, to bridge theory and experiment.

\section{Experimental Validation}
\label{sec_experimental}

\begin{figure*}[th!]
    \centering
    \includegraphics[width=\textwidth]{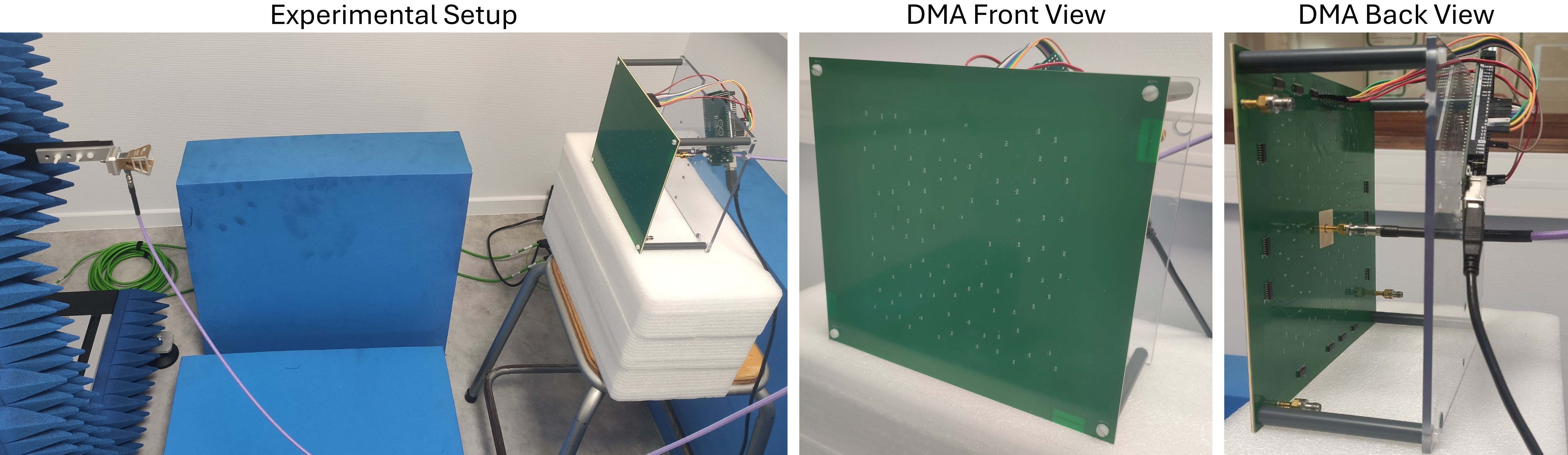}
    \caption{Photographic images of the experimental setup and close-up views of front and back of the DMA. The DMA is fed by a coaxial cable (purple) and configured via the Arduino which receives logic commands via a USB cable (black). The other three ports seen on the DMA's back are unused positioning fiducials that are not connected to the cavity and not used in this paper. The 96 meta-atoms are seen on the DMA's front. A horn antenna facing the DMA is mounted on a 2D positioning stage to measure the field radiated by the DMA.}
    \label{FigX}
\end{figure*}

In this section, we  demonstrate \textit{experimentally} that an increase in the DMA MC strength results in both a stronger DMA radiation pattern sensitivity and a stronger non-linearity of the DMA radiation pattern parametrization, as well as a greater ability to tailor the DMA radiation pattern to the needs of a wireless functionality. Our objective in this section is to demonstrate a qualitative agreement between our experimental observations and our theoretical and numerical results from Sec.~\ref{sec_Theory} and Sec.~\ref{sec_numerical}, respectively. Establishing a quantitative link between the experiment and our theoretical and numerical results is deferred to future work because it requires the development of techniques to estimate the model parameters for a given experimental DMA prototype, which is beyond the scope of this paper. 

Our experiments are based on a single-feed chaotic-cavity-backed DMA similar to the one proposed in~\cite{sleasman2020implementation} (details below). Fabricating tens of distinct DMA prototypes that differ in the cavity’s lossiness or leakiness to systematically explore different DMA MC strengths is infeasible. Instead, we introduce an original time-gating technique that enables post-processing adjustments to the DMA's effective MC strength. Time gating is an established post-processing technique for suppressing scattered waves in imaging applications~\cite{7362172,barolle2021manifestation,9926204} and antenna characterization~\cite{1296154,timegating_XLIM}. The fundamental insight is that the delay time of a path is directly related to the number of scattering events along the path -- in line with the multi-bounce picture of wave propagation in the DMA. Wave energy exiting the DMA after longer delay times has undergone more bounces between the meta-atoms. If there was no MC in the DMA, there would be no wave energy exiting the DMA after long delay times. Hence, by  considering only the wave energy radiated by the DMA up to some delay time $\tau$, we can suppress the contributions of paths associated with delay times beyond $\tau$, thereby effectively reducing the DMA's effective MC strength. In other words, we have continuous control in post-processing over the DMA's MC strength by selecting $\tau$. Of course, the MC strength in the physical prototype is an upper limit for the values that we can explore (and corresponds to $\tau\rightarrow \infty$).

We conduct our experiments in a model-free manner. We do believe that the parameters of the compact DMA system model from Sec.~\ref{sec_system_model} can be calibrated to accurately describe a given DMA prototype using techniques similar to the ones developed for RIS in~\cite{sol2023experimentally,del2024minimal,V2NA_2p0,del2024physics}. However, transposing these existing techniques to DMAs is beyond the scope of this paper, not least because such a transposition is not required to (approximately) measure the quantities of interest here.

\subsection{Experimental Details}

\subsubsection{DMA Design} Our DMA prototype is displayed in Fig.~\ref{FigX}. The DMA consists of a 4-layer PCB based on the design from~\cite{sleasman2020implementation}.
The two upper conducting layers are separated by a low-loss 1.52~mm thick Rogers 4003 substrate. The combination of the two upper conducting layers with a fence of 841 vias (0.2~mm diameter) creates a quasi-2D cavity. The shape of the via fence is irregular, giving rise to wave chaos in the cavity, and spans an area of roughly $15\times 15\  \mathrm{cm}^2$. This chaotic cavity is excited using a central coaxial feed (3811-40092, Amphenol). 96 programmable meta-atoms are embedded into the upper conducting layer, at random positions within the cavity (up to a constraint regarding their minimal separation). The meta-atom design is based on~\cite{yoo2016efficient}. Each meta-atom contains a PIN diode (MADP000907-14020W) whose bias voltage can be controlled individually. Specifically, binary bias voltage control is enabled by a DC bias circuitry embedded between the second, third and fourth PCB layers that are separated by 0.4~mm thick FR4 substrates. Each PIN diode is in series with a $750\ \Omega$ resistor to limit the current. The 96 bias voltages are controlled by twelve eight-bit shift registers (SN74HC595N, Texas Instruments) in series. The configurations of the shift registers and the power supply are provided by an Arduino development board. To reconfigure the DMA, a 96-element binary vector $\mathbf{b}\in\mathbb{B}^{96}$ is sent from Python via a serial connection to the Arduino, which divides the vector into twelve vectors of eight bits and shifts the registers accordingly. Note that we do not know the mapping from $\mathbf{b}$ to $\mathbf{c}\in\mathbb{C}^{96}$ for our DMA prototype.

\subsubsection{Experimental Setup} We measure the field radiated by the DMA with a broadband double-ridged horn antenna (DRH67) that is mounted on a two-axis linear translation stage. The horn antenna faces the DMA, as seen in Fig.~\ref{FigX}. Our choice for the horn antenna is motivated by its low-dispersion broadband matching because our time-gating technique requires an acquisition of the DMA's radiated field across a broad spectrum, ideally without dispersive distortions from the receiving antenna. Of course, the field captured by the horn antenna must be understood as a weighted integral of the field over parts of the scene rather than the field at a single point, but this does not pose an issue for our present study. Both the DMA and the horn antenna are connected via coaxial cables to a vector network analyzer (Rohde \& Schwarz ZVA67) that measures the transmission coefficient between the two antenna ports for 16,001 linearly spaced frequency points between 15~GHz and 26.5~GHz.\footnote{This wideband measurement is required for our time-gating analysis to suppress multipath contributions beyond a desired delay time at the operation frequency. To be clear, our optimization targets a \textit{single} frequency of operation, and no claim of broadband DMA operation across 11.5 GHz is implied.} We denote this transmission coefficient by $h(f,\mathbf{b})$ because it is the wireless channel between the two antennas' ports, parametrized by the DMA configuration $\mathbf{b}$.

\subsubsection{Measurement Procedure} Algorithm~\ref{exp_meas_procedure} summarizes our measurement procedure. For ten random positions of the horn antenna (within a $30\times 30\ \mathrm{cm}^2$ area that is in front of, parallel to, and centered with respect to the DMA's surface), we acquire the data necessary to evaluate our linearity metric $\zeta$ as well as the normalized Jacobian and Hessian of the radiated field at any frequency and for any value of $\tau$; in addition, at each of these horn antenna positions, we perform coordinate ascent optimizations using Algorithm~\ref{Alg_CoordAsc} to obtain the achievable channel gain enhancement for all combinations of $\tau \in [2.5,2.75,3,3.25,3.5,4,4.5]\ \mathrm{ns}$ and $f_0 \in [18,19,20,21]\ \mathrm{GHz}$. The chosen values of $\tau$ are based on the line-of-sight (LOS) time delay of roughly 3~ns between the two antennas (measured from antenna port to antenna port, i.e., including propagation in the antennas). The chosen values of $f_0$ are equally spaced within the DMA's operation bandwidth. Note that each combination of $\tau$ and $f_0$ requires a separate run of the coordinate ascent method.

\begin{algorithm}[t]
\footnotesize
\label{exp_meas_procedure}
\For{\textrm{each considered horn antenna position}} 
{
    \For{$0 < r < 512$}
    {
        Measure $h(f,\mathbf{b}^{0,r})$, where $\mathbf{b}^{0,r}$ is a random binary DMA configuration.\\
        \For{$0 \leq i < 96$}
        {
            Measure $h(f,\mathbf{b}^{i,r})$, where $\mathbf{b}^{i,r}$ equals $\mathbf{b}^{0,r}$ except that the $i$th entry is flipped.\\
            \For{$i < j < 96$}
            {
                Measure $h(f,\mathbf{b}^{ij,r})$, where $\mathbf{b}^{ij,r}$ equals $\mathbf{b}^{0,r}$ except that the $i$th and $j$th entries are flipped.
            }
        }
    }
    \For{\textrm{each considered value of} $\tau$}
    {
        Calculate the time-gated versions $h^\mathrm{TG}(f,\mathbf{b}^{0,r},\tau)$ of all previously measured $h(f,\mathbf{b}^{0,r})$.\\
        \For{\textrm{each considered target frequency} $f_0$}
    {
        Run Algorithm~\ref{Alg_CoordAsc}.
    }
    }
}
\caption{Experimental Measurement Procedure.}
\end{algorithm}

\subsection{Experimentally Evaluated Metrics}

As mentioned, we experimentally evaluate only model-free metrics. 
Given the 1-bit programmability of our prototype's meta-atoms, we determine finite-difference approximations (denoted by a bar) of the Jacobian and Hessian of $h(f,\mathbf{b})$ with respect to $\mathbf{b}$ based on the measurements from the first part of Algorithm~\ref{exp_meas_procedure}:
\begin{subequations}\label{eq25}
    \begin{equation}
        \bar{\mathbf{j}}_i(f,\mathbf{b}^{0,r}) = h(f,\mathbf{b}^{i,r}) - h(f,\mathbf{b}^{0,r}),
    \end{equation}
        \begin{equation}
        \bar{\mathbf{H}}_{ij}(f,\mathbf{b}^{0,r}) = h(f,\mathbf{b}^{ij,r}) - h(f,\mathbf{b}^{i,r}) - h(f,\mathbf{b}^{j,r}) + h(f,\mathbf{b}^{0,r}).
        \label{eq25b}
    \end{equation}
\end{subequations}
\noindent Note that we can only evaluate off-diagonal entries of $\bar{\mathbf{H}}_{ij}(f,\mathbf{b}^{0,r})$ due to the meta-atoms' binary programmability. 
Based on the time-gated versions of the measured channels, we compute the approximate time-gated Jacobian $\tilde{\mathbf{j}}^\mathrm{TG}(f,\mathbf{b}^{0,r},\tau)$ and Hessian $\tilde{\mathbf{H}}^\mathrm{TG}(f,\mathbf{b}^{0,r},\tau)$ similar to Eq.~(\ref{eq25}). We then compute $\langle\sigma\rangle$, $\langle\theta\rangle$, $\langle\zeta\rangle$ and $\langle\eta\rangle$ analogous to Sec.~\ref{sec_metrics_numerics}.

\subsection{Experimental Results}

\begin{figure}
    \centering
    \includegraphics[width=0.72\columnwidth]{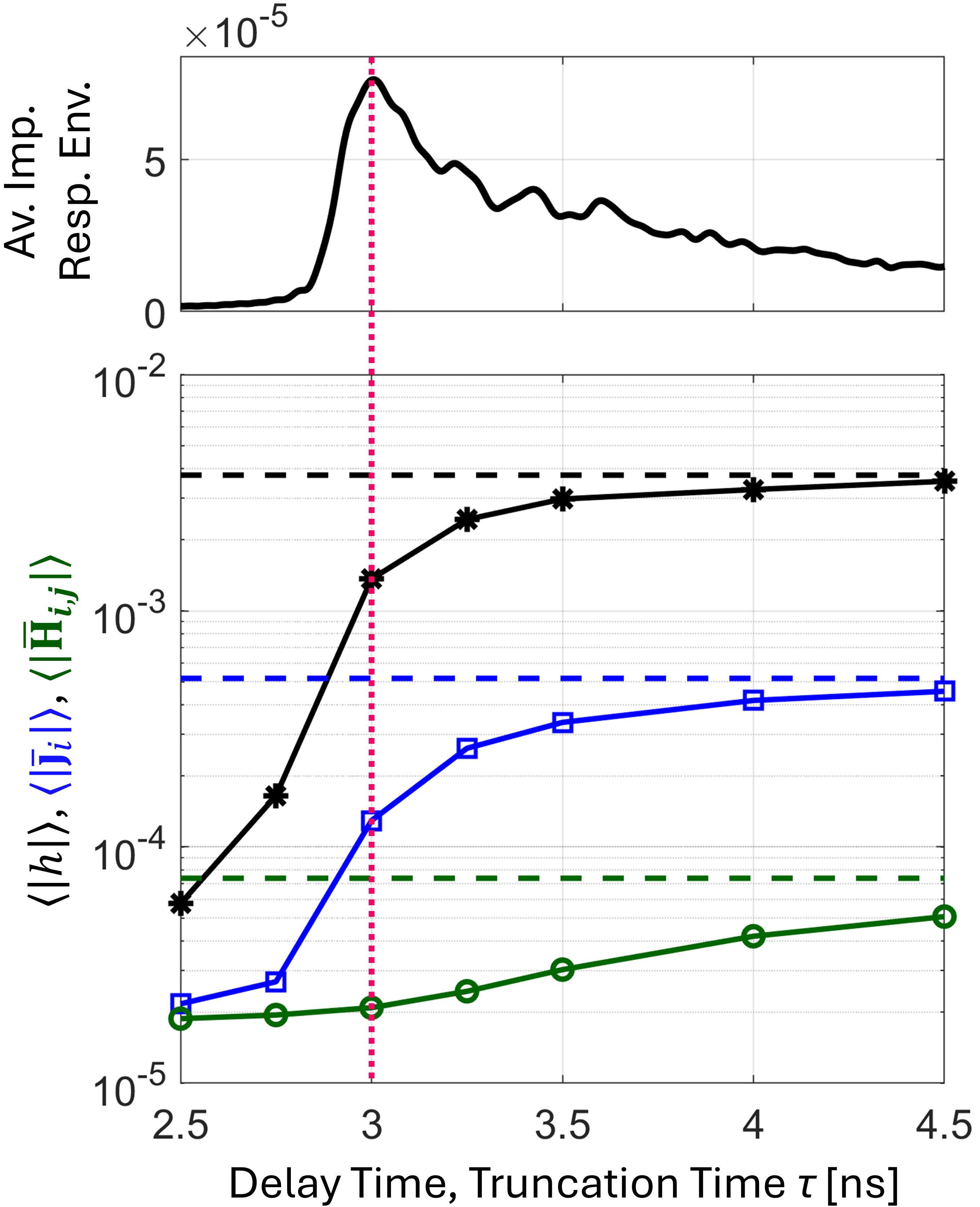}
    \caption{The top row displays the average impulse response envelope of the wireless channel as a function of the delay time. The vertical dotted pink line indicates the LOS delay time. The bottom row displays the experimentally measured average magnitudes of the wireless channel (black), the entries of its finite-difference Jacobian (blue) and the off-diagonal entries of its finite-difference Hessian (green), as a function of the time-gating truncation time $\tau$. In the limit $\tau\rightarrow\infty$, the curves converge to the horizontal dashed lines obtained without time gating.}
    \label{FigYYY}
\end{figure}

\begin{figure}
    \centering
    \includegraphics[width=\columnwidth]{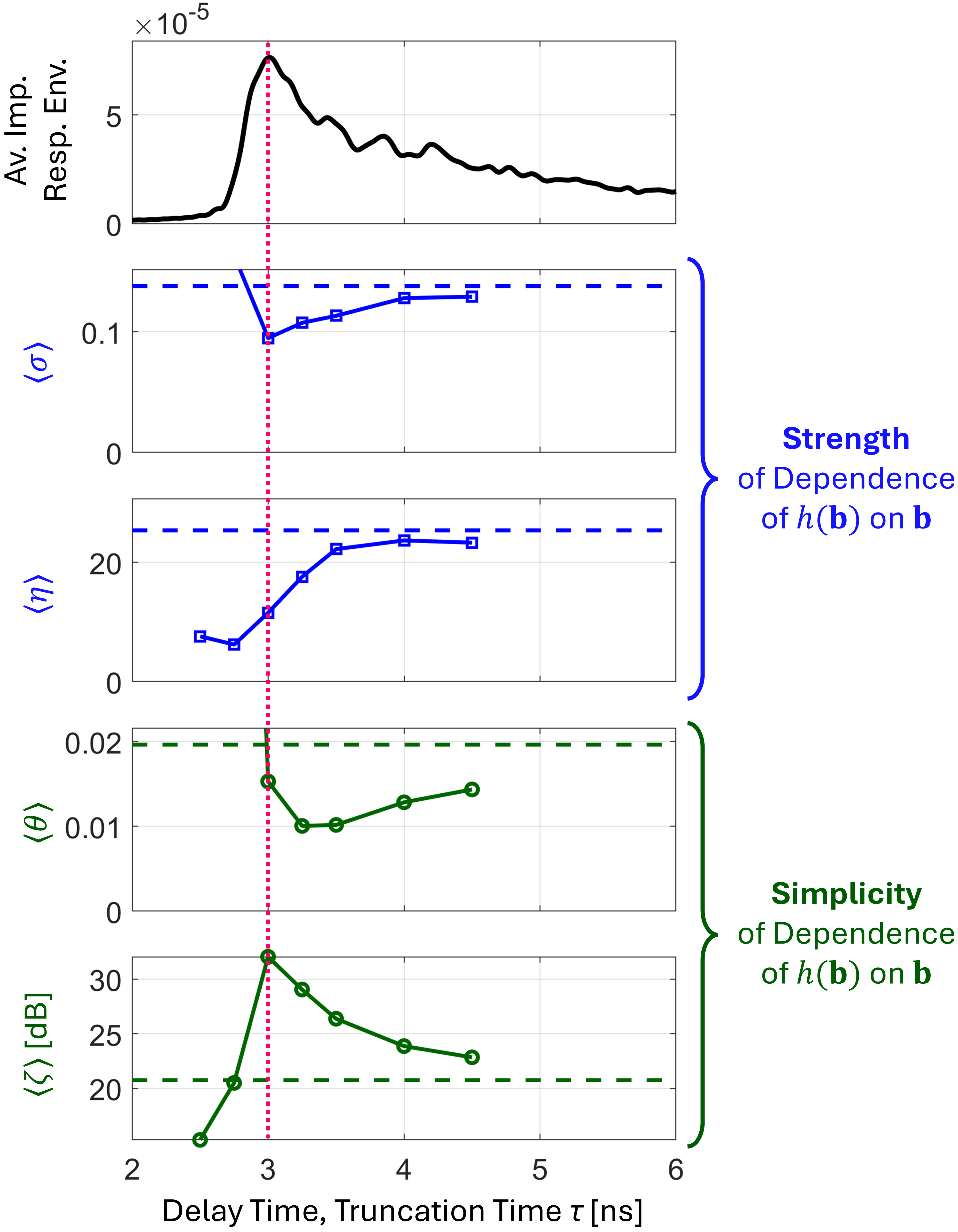}
    \caption{Experimental results for the trade-off between strength and simplicity of the dependence of $h(\mathbf{b})$ on $\mathbf{b}$, governed by the DMA's MC strength. We use the time-gating truncation time $\tau$ to control the DMA's effective MC strength in post-processing. The larger $\tau$ is, the stronger is the effective MC strength (see main text for detailed explanations).     
    The top row reproduces for reference the average impulse response envelope, and the pink vertical line indicates again the LOS delay time. The subsequent rows display the average sensitivity $\langle\sigma\rangle$, the average channel gain enhancement $\langle\eta\rangle$, the average normalized off-diagonal Hessian entries $\langle\theta\rangle$, and the average linearity metric $\langle\zeta\rangle$, for different time-gating truncation times $\tau$. In the limit $\tau\rightarrow\infty$, the curves converge to the horizontal dashed lines obtained without time-gating.}
    \label{FigYY}
\end{figure}

We first inspect the  average \textit{absolute} magnitudes of the wireless channel, the entries of the Jacobian and the off-diagonal entries of the Hessian in Fig.~\ref{FigYYY} as a function of $\tau$. It is important to interpret the $\tau$ dependence in light of the average channel impulse response envelope shown on top in Fig.~\ref{FigYYY}. The LOS delay of roughly 3~ns makes sense in sight of the separation of the two antennas and the propagation inside the antennas, and it implies that no signal can be received for $\tau<3\ \mathrm{ns}$ (except for the build-up of the finite-bandwidth LOS tap). Results for $\tau<3\ \mathrm{ns}$ can hence not be interpreted meaningfully, and are measured with a poor signal-to-noise ratio (the wireless channel magnitude is an order of magnitude lower at $\tau=2.75\ \mathrm{ns}$ than at $\tau=3\ \mathrm{ns}$). It is apparent in Fig.~\ref{FigYYY} that the Jacobian and Hessian entries are, respectively, one and two orders of magnitude weaker than the wireless channel. In the limit $\tau\rightarrow\infty$, all three quantities converge to their values without time gating that are indicated by horizontal dashed lines. The fact that the wireless channel's absolute magnitude varies with $\tau$ emphasizes the importance of studying the normalized Jacobian and Hessian to make a fair comparison with fixed radiation efficiency, as we do in Fig.~\ref{FigYY}. Upon careful visual inspection, it is apparent in Fig.~\ref{FigYYY} that the wireless channel converges quicker to the dashed reference than the Jacobian, which in turn converges much quicker than the Hessian. This observation reveals that the \textit{relative} DMA radiation pattern sensitivity increases for larger $\tau$ (and, hence, for larger effective DMA MC strengths), and that the \textit{relative} non-linearity of the DMA radiation pattern parametrization increases even stronger for larger $\tau$. These trends confirm that the MC strength governs the trade-off between strength and simplicity of the dependence of the radiation pattern on the DMA configuration.

To reveal these trends more directly, we plot the average \textit{relative} magnitudes of the Jacobian, $\langle\sigma\rangle$, and Hessian, $\langle\theta\rangle$, in Fig.~\ref{FigYY}. The average relative sensitivity without time gating (i.e., for $\tau\rightarrow\infty$) is 45.4~\% stronger than for time-gating at the LOS delay $\tau=3\ \mathrm{ns}$. In other words, stronger MC in the DMA results in a substantially larger sensitivity of the radiated field to the DMA configuration. This larger sensitivity translates into massive benefits for wireless engineers to tailor the radiation pattern. Indeed, for the prototypical problem of enhancing the UE channel gain, we observe that the 45.4~\% improvement in relative sensitivity translate into 219.0~\% improvement in the channel gain enhancement, which rises from 11.6 at $\tau=3\ \mathrm{ns}$ to 25.4 for $\tau\rightarrow\infty$. The flipside of this benefit of stronger MC is the increased non-linearity of the radiation pattern parametrization. Indeed, the linearity metric $\langle \zeta \rangle$ drops from 32.1~dB at $\tau=3\ \mathrm{ns}$ to 20.7~dB for $\tau\rightarrow\infty$. Correspondingly the average relative magnitude of the off-diagonal entries of the Hessian double from their minimum at $\tau = 3.25$~ns to the limit of $\tau\rightarrow\infty$. As explained earlier, the values for $\tau<3$~ns cannot be meaningfully interpreted because they correspond to unphysical delay times. In the case of the Hessian, even the value at $\tau=3$~ns appears abnormal, and only from $\tau=3.25$~ns onwards the trend is clear. This may be due to the fact that the Hessian originates from trajectories within the DMA that have encountered at least two meta-atoms and correspond hence to delay times that are necessarily larger than the LOS delay time of 3~ns. In other words, for the LOS delay time of 3~ns we cannot expect any meaningful measurement of the Hessian.

\subsection{Summary}
In this section, we introduced a time-gating technique to control the effective MC strength in post-processing. We then leveraged this technique to report experimental evidence based on a DMA prototype in the lower part of the K-band. We experimentally measured the Jacobian and Hessian of the radiation pattern. We experimentally validated the trends seen in previous sections, namely that a stronger MC strength results in a larger sensitivity (which translates into substantial improvements in channel gain enhancement) and a larger Hessian (which translates into a lower linearity metric).

\section{Discussion}
\label{sec_Discussion}

The goal of this paper is to evidence that strong MC enhances the sensitivity of the radiation pattern to the DMA configuration, and that an enhanced sensitivity improves the ability to tailor the radiation pattern to a desired functionality. Consequently, we argue that strong MC should be embraced to maximize the radiation pattern tunability with a given number of tunable elements. In contrast to our argument, the literature contains some works that actively seek to limit (rather than to maximize) the sensitivity~\cite{xu2022extreme,xu2022wide}. To the best of our understanding, there is no evidence of the connection between MC strength, sensitivity and tunability in these works; instead, their motivation to limit the sensitivity originates from considerations of vulnerability to fabrication inaccuracy and operation bandwidth. In this discussion, we address these two concerns. 

\subsection{Vulnerability to fabrication inaccuracies}

Most schemes for operating reconfigurable antennas in experiments (including~\cite{xu2022wide}) rely on a forward model of the antenna whose parameters are known analytically or obtained in numerical simulations based on the known antenna design. Naturally, such schemes are vulnerable to fabrication inaccuracies. The more sensitive the radiation pattern is to the DMA configuration, the more sensitive it is also to inaccuracies in the model parameters. In such schemes, the vulnerability to fabrication inaccuracies raised in~\cite{xu2022extreme,xu2022wide} is thus a valid concern.

Our vision for operating reconfigurable antennas in experiments is different. We envision that a compact model is calibrated in situ, i.e., its parameters are estimated for an experimentally given, fabricated device. Thereby, the concern about vulnerability to fabrication inaccuracies inherently vanishes. We are confident that this approach is feasible given our recent works on the conceptually related problem of estimating model parameters for RIS-parametrized wireless channels~\cite{sol2023experimentally,del2024minimal,V2NA_2p0,del2024physics,del2025experimental,del2025ambiguity}. For the purpose of experimental parameter estimation, the model should be as compact as possible to avoid over-parametrization resulting in avoidable ambiguities and avoidable computational complexity. Our model formulation in Sec.~\ref{sec_system_model} is maximally compact because it collapses architecture-specific details of the coupling mechanisms between feeds and meta-atoms into a background coupling operator; in any case, the details of these coupling mechanisms could not be identified unambiguously based on experimental measurements.
However, the algorithmic developments to transpose the parameter estimation approaches from~\cite{sol2023experimentally,del2024minimal,V2NA_2p0,del2024physics,del2025experimental,del2025ambiguity} in the realm of RIS to DMAs are beyond the scope of this paper and are thus deferred to future work. We emphasize that all required experimental results for this paper are obtained in a model-agnostic manner, so that this paper did not require the development of parameter-estimation algorithms.

\subsection{Operation bandwidth}

\begin{figure}[b]
    \centering
    \includegraphics[width=\columnwidth]{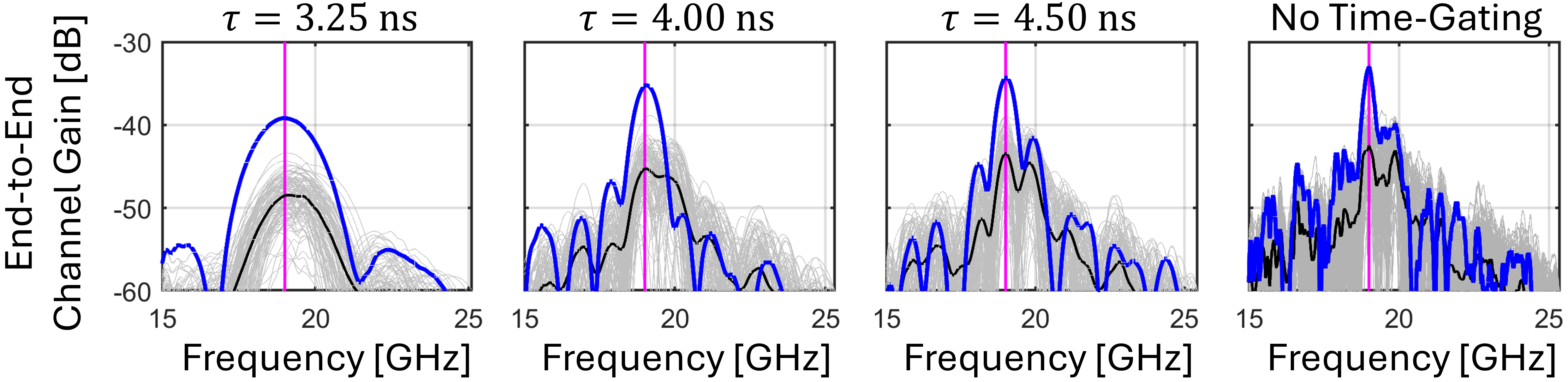}
    \caption{Selected examples of end-to-end channel spectra for $f_0=19$~GHz for four different choices of $\tau$. (blue: optimized; gray: random; black: average over random) }
    \label{Fig8}
\end{figure}

Because of the fundamental connection between dwell time and sensitivity~\cite{del2021deeply}, an enhanced sensitivity is also expected to result in a more rapid spectral decorrelation of the radiation pattern. In other words, if the DMA configuration is optimized to achieve a desired functionality at a single frequency, the interval of frequencies around the optimized frequency for which the desired functionality is also roughly achieved is narrower. This effect can indeed be observed in our experimental data, as documented for selected examples in Fig.~\ref{Fig8} and summarized on average over all examples in Table~\ref{tab:tau_bw}. The larger the truncation time $\tau$ is (which is our proxy for the MC strength), the narrower is the peak around the targeted frequency in the optimized spectrum of the end-to-end channel gain.

\begin{table}[b]
\centering
\caption{Experimentally evaluated FWHM bandwidth of the optimized end-to-end channel, averaged over individual optimizations for each receiver position and operating frequency.}
\label{tab:tau_bw}
\begin{tabular}{c|cccccccc}
\midrule
$\tau$~[ns] & 2.5 & 2.75 & 3.00 & 3.25 & 3.50 & 4.00 & 4.50 & $\infty$ \\
\midrule
BW~[GHz] & N/A & N/A & 4.00 & 2.14 & 1.46 & 0.80 & 0.58 & 0.35 \\
\bottomrule
\end{tabular}
\end{table}

However, even for the case of $\tau\rightarrow\infty$, we still have an average peak width of 0.35~GHz which can comfortably accommodate at least one wireless link. 
We emphasize that the relatively narrow bandwidth does not preclude operations of our DMA across a much wider frequency range. On the one hand, we can sequentially switch between DMA configurations optimized for different central operating frequencies. On the other hand, we can optimize the DMA configuration to simultaneously serve multiple carrier frequencies (by trading off channel gain with bandwidth).

Finally, we note that the spectral selectivity associated with strong MC can actually be a desirable feature enabling a frugal use of spectral resources as well as interference rejection. In addition, applications like computational imaging and wireless power transfer often target single-frequency operation to alleviate the cost of the required RF chains.

Altogether, while there is a clear trade-off between MC strength and operation bandwidth, we have shown that the largest conceivable MC strength in our DMA prototype still offers sufficient bandwidth for a typical wireless link. While our optimizations in the present paper were focused on single-frequency optimizations, future work will explore the optimization of the DMA configuration for desired functionalities across multiple frequency points.

\section{Conclusion}
\label{sec_Conclusion}

To summarize, we have demonstrated that a DMA's MC strength governs a fundamental trade-off between strength and simplicity of the dependence of the DMA's radiation pattern on the DMA's configuration, based on systematic theoretical, numerical and experimental evidence. On the one hand, the stronger the MC between the meta-atoms is, the more sensitive the radiation pattern is to the DMA configuration. We demonstrated that this enhanced sensitivity translates into substantial improvements in our ability to optimize the DMA configuration for a desired wireless functionality, considering channel gain enhancement as prototypical problem. On the other hand, stronger MC makes the mapping from configuration to radiation pattern more non-linear, complicating the accurate modeling and physics-compliant optimization of DMAs for wireless communications.

Our work unveils untapped potential for substantially improved performance in DMA-assisted wireless communications that was overlooked so far because most existing studies on DMAs neglected or deliberately mitigated MC. The price for reaping the benefits of strong MC between meta-atoms is the requirement for physics-compliant modeling and optimization. We presented a compact and implementation-agnostic physics-compliant DMA model and are confident that its parameters can be estimated to describe any given DMA prototype, using calibration techniques similar to those already experimentally validated in the RIS context~\cite{sol2023experimentally,del2024minimal,V2NA_2p0,del2024physics,del2025experimental,del2025ambiguity}.

Looking forward, our findings may encourage wireless practitioners to fundamentally reconsider the current assessment of MC in DMAs as a vexing nuance. Concretely, we expect DMA architectures that favor strong MC such as the chaotic-cavity-backed ones considered in our studies to gain traction, and that efficient DMA optimization algorithms capable of coping with strong MC will be developed. We also mentioned an idea for a new generation of ``beyond-diagonal'' DMA (BD-DMA) hardware with tunable MC which can enable attaining the upper bound on the radiation pattern sensitivity; an initial exploration of performance gains with BD-DMAs can be found in our recent follow-up work~\cite{prod2025beyond}. 

We expect that the unveiled fundamental dependence of the ability to control the transfer function of a reconfigurable wave system on the MC strength generalizes to ESPAR antennas~\cite{harrington1978reactively,schlub2003seven,sun2004fast,kawakami2005electrically,lu2005dielectric,luther2012microstrip,movahedinia2018}, reconfigurable pixel-based antennas~\cite{Flaviis_PixelAntenna,rodrigo2012frequency,MURCH_TAP_PixelAntenna}, RIS~\cite{nerini2024global,semmler2024decoupling,liu2025optimization}, wireless localization~\cite{del2021deeply}, and wave-domain physical neural networks~\cite{momeni2023backpropagation}.

\appendices

\section{Derivation of Eq.~(\ref{eq1000a})}
\label{AppendixA}

Our starting point to derive Eq.~(\ref{eq1000a}) is the standard matrix identity $\frac{\partial \mathbf{A}^{-1}}{\partial x} = -\mathbf{A}^{-1} \frac{\partial \mathbf{A}}{\partial x}   \mathbf{A}^{-1} $ (see Eq.~(59) in~\cite{petersen2008matrix}). Applied to our problem, the identity yields
\begin{equation}
    \frac{\partial \mathbf{W}^{-1}}{\partial c_i} = - \mathbf{W}^{-1} \frac{\partial \mathbf{W}}{\partial c_i} \mathbf{W}^{-1}.
     \label{eqA11}
\end{equation}
It follows directly from Eq.~(\ref{eq5}) that
\begin{equation}
   \frac{\partial \mathbf{W}}{\partial c_i} = \begin{bmatrix} 
	\mathbf{0}_{\mathcal{FF}} & \mathbf{0}_{\mathcal{FM}} \\
	\mathbf{0}_{\mathcal{MF}} & -\mathbf{E}_{ii}\\
	\end{bmatrix},
     \label{eqA12}
\end{equation}
where $\mathbf{E}_{ii}$ denotes an $N_\mathrm{M} \times N_\mathrm{M}$ matrix whose entries are zero except for the $i$th diagonal entry which is unity. Substituting Eq.~(\ref{eqA12}) into Eq.~(\ref{eqA11}) and selecting the $\mathcal{MF}$ block yields
\begin{equation}
\begin{split}
\mathbf{j}^\mathrm{p}_i = \left[ \frac{\partial \mathbf{W}^{-1}}{\partial c_i} \right]_\mathcal{MF} e^\mathrm{inc} &= \left[ \mathbf{W}^{-1}\right]_\mathcal{MM}  \mathbf{E}_{ii}\left[\mathbf{W}^{-1}\right]_\mathcal{MF} e^\mathrm{inc}\\ &= \mathbf{\Omega}_{\mathcal{M}\mathcal{M}_i}(\mathbf{c}) \mathbf{\Omega}_{\mathcal{M}_i\mathcal{F}}(\mathbf{c})e^\mathrm{inc}\\ &= \mathbf{\Omega}_{\mathcal{M}\mathcal{M}_i}(\mathbf{c}) \mathbf{p}_{\mathcal{M}_i}(\mathbf{c}).
\end{split}
\end{equation}

\section{Derivation of Eq.~(\ref{eq_bound})}
\label{AppendixB}
We start by separating $\tilde{\mathbf{W}} - \mathrm{diag}(\mathbf{c})$ into its diagonal part $\mathbf{A}\triangleq \text{diag}(\text{diag}(\tilde{\mathbf{W}} - \mathrm{diag}(\mathbf{c})))$ and its off-diagonal part $\mathbf{B} \triangleq \tilde{\mathbf{W}} - \mathrm{diag}(\mathbf{c})-\mathbf{A}$, so that
\begin{equation}
    \mathbf{\Omega}_\mathcal{MM}(\mathbf{c}) =\big( \tilde{\mathbf{W}} - \mathrm{diag}(\mathbf{c})\big)^{-1} = \big(  \mathbf{A} + \mathbf{B} \big)^{-1}.
\end{equation}
We denote the singular values of $\mathbf{A}$ by $\sigma_1 \geq \dots \geq \sigma_{N_\mathrm{M}}$ (they are just the sorted moduli of the diagonal entries of $\mathbf{A}$). We further denote by $\tilde{\sigma}_1 \geq \dots \geq \tilde{\sigma}_{N_\mathrm{M}}$ the singular values of $\mathbf{A} + \mathbf{B}$. It follows immediately that $\|\mathbf{\Omega}_\mathcal{MM}(\mathbf{c})\|_2  = \frac{1}{\tilde{\sigma}_{N_\mathrm{M}}}$. 

Using Weyl's inequality for singular values \cite{Weyl1912} \cite[Theorem 1]{stewart1998perturbation} yields $
|\sigma_{N_\mathrm{M}} - \tilde{\sigma}_{N_\mathrm{M}}| \leq \|\mathbf{B}\|_2$. Then, using the reverse triangle inequality, we bound $\tilde{\sigma}_{N_\mathrm{M}}$ as $\tilde{\sigma}_{N_\mathrm{M}} \geq \sigma_{N_\mathrm{M}} -  \|\mathbf{B}\|_2$, provided that $\sigma_{N_\mathrm{M}} \geq 
 \|\mathbf{B}\|_2$ because otherwise we would have the trivial inequality $\tilde{\sigma}_{N_\mathrm{M}} \geq 0$. 
Finally, defining $\tilde{\mu} \triangleq  \|\mathbf{B}\|_2.\|\mathbf{A}^{-1}\|_2 = \frac{\|\mathbf{B}\|_2}{\sigma_{N_\mathrm{M}}}$ and inverting both sides of the previous inequality yields
\begin{equation}
    \|\mathbf{\Omega}_\mathcal{MM}(\mathbf{c})\|_2 \leq \frac{1}{\sigma_{N_\mathrm{M}}(1-\tilde{\mu})},
\end{equation}
which is Eq.~(\ref{eq_bound}).
Note that $\tilde{\mu} = \mu_\mathrm{n} \frac{\sigma_1}{\sigma_{N_\mathrm{M}}} = \mu_\mathrm{n} \kappa(\mathbf{A})$, where $\kappa(\mathbf{A})$ denotes the condition number of $\mathbf{A}$.

The bound would be attained with equality if and only if the singular vector associated with the greatest singular value of $\mathbf{B}$ were perfectly collinear with the singular vector associated with the smallest singular value of $\mathbf{A}$, differing only in sign. Realizing this alignment would require perfect control over the off-diagonal part of $\tilde{\mathbf{W}}$, which is not the case for the DMA considered in this paper.

\bibliographystyle{IEEEtran}


\end{document}